\documentclass[hyper]{JHEP3}

\usepackage{amssymb,amsthm}

\newcommand{\C}{{\mathbb C}}

\newcommand{\cL}{{\mathcal L}}
\newcommand{\cH}{{\mathcal H}}

\newcommand{\cO}{{\mathcal O}}

\newcommand{\cD}{{\mathcal D}}

\newcommand{\cS}{{\mathcal S}}

\newcommand{\beq}{\begin{equation}}
\newcommand{\eeq}{\end{equation}}

\newcommand{\ber}{\begin{eqnarray}}
\newcommand{\eer}{\end{eqnarray}}

\def\im{{\rm i}}
\def\GR{{\rm GR}}
\def\eff{{\rm eff}}

\def\eq{{\rm eq}}

\def\etal{{et al.}}

\title{Cosmological perturbations in a family of deformations of general relativity}

\author{Kirill Krasnov\\ School of Mathematical Sciences, University of Nottingham, Nottingham
NG7 2RD, UK \\ E-mail: \email{kirill.krasnov@nottingham.ac.uk}}
\author{Yuri Shtanov\\ Bogolyubov Institute for Theoretical Physics, Kiev 03680, Ukraine\\
Faculty of Physics, Taras Shevchenko National University of Kiev, Ukraine\\
E-mail: \email{shtanov@bitp.kiev.ua}}

\abstract{We study linear cosmological perturbations in a previously introduced family of
deformations of general relativity characterized by the absence of new degrees of
freedom. The homogeneous and isotropic background in this class of theories is unmodified
and is described by the usual Friedmann equations. The theory of cosmological
perturbations is modified and the relevant deformation parameter has the dimension of length.
Gravitational perturbations of the scalar type can be described by a certain
relativistic potential related to the matter perturbations just as in
general relativity. A system of differential equations describing the evolution of this potential
and of the stress-energy density perturbations is obtained. We find that the
evolution of scalar perturbations proceeds with a modified effective time-dependent speed
of sound, which, contrary to the case of general relativity, does not vanish even at the
matter-dominated stage.  In a broad range of values of the length parameter controlling
the deformation, a specific transition from the regime of modified gravity to the regime
of general relativity in the evolution of scalar perturbations takes place during the
radiation domination. In this case, the resulting power spectrum of perturbations in
radiation and dark matter is suppressed on the comoving spatial scales that enter the
Hubble radius before this transition. We estimate the bounds on the deformation parameter for
which this suppression does not lead to observable consequences. Evolution of scalar perturbations
at the inflationary stage is modified but very slightly and the primordial spectrum generated
during inflation is not noticeably different from the one
obtained in general relativity.}


\begin{document}

\section{Introduction}

The future astrophysical and cosmological data will, among other things, allow one to
perform deeper tests of our theory of gravity, general relativity (GR)\@. Although there
are no compelling reasons to question seriously the validity of GR, several puzzles
associated with its long-distance behavior, primarily those connected with the phenomena
of dark energy and dark matter, make it an important task to see whether GR continues to
be as good a description of gravity on astrophysical and cosmological scales as it is on
the scales of Solar System. Modern literature is abundant with various proposals of
modified gravity addressing this issue. A common feature of almost all such modifications
is that new propagating degrees of freedom are introduced in the gravitational sector, in
the form of additional fields or higher derivatives of the metric field. A typical
example is $f(R)$ gravity, which in its original variables is a purely metric theory with
higher derivatives but also can be transformed into a scalar-tensor theory.  Extensions
of this kind typically change both the Friedmann equations describing the evolution of a
homogeneous and isotropic universe and the equations describing cosmological
perturbations, by which modifications they can be tested.

It is, perhaps, less known that one can modify GR without introducing any new propagating
degrees of freedom. Modification of this kind, in particular, should be expected not to
change the homogeneous background evolution in cosmology, affecting only the way the
perturbations evolve. Some of the generic features of this scheme were explored in
\cite{Skordis:2008vt} without specifying a concrete theory.

A specific class of modifications of GR without new propagating degrees of freedom was
proposed by one of us in \cite{Krasnov:2006du}, with the same class of theories envisaged
earlier in a different formulation in \cite{Bengtsson:1990qg}. These theories
legitimately can be called deformations of GR in the sense that there exist continuous
paths in the parameter space smoothly connecting any of them with GR.  The theories in
this class provide a new framework for testing general relativity, since, using the
observational data, one can determine how close is the real-world gravity to GR within a
class of its deformations. In one of our
previous papers \cite{Krasnov:2007ky}, we were concerned with the spherically symmetric
solution in this class of theories, its features and possible implications. In the
present paper, we turn to the description of cosmology.  Our aim is to set a stage for
possible tests of the theories of \cite{Krasnov:2006du} by working out the modified
equations describing the evolution of cosmological perturbations.

Before describing the class of theories under investigation in detail, it is worth discussing their
basic properties in elementary terms. If viewed as purely metric theories, they can be regarded as
the usual effective theories with a nonlocal Lagrangian that can be expanded into an infinite
series of local terms built from the curvature invariants: ${\mathcal L} =
\alpha_1 R + \alpha_2 R^2 + \alpha_3 R_{\mu\nu} R^{\mu\nu} + \alpha_4
R_{\mu\nu\sigma\tau} R^{\mu\nu\sigma\tau} + \ldots\,$ The coefficients in front of these
invariants, however, are not arbitrary in the present case, but are such that the whole infinite series
can be resummed with the help of auxiliary non-propagating fields. In the end, one obtains a local
second-derivative theory with only two propagating degrees of freedom (two polarizations
of the gravitational wave), and a certain limit can be taken to recover GR if one so wishes, see
\cite{Krasnov:2009ik}.

To elucidate this idea, we illustrate it on a simple example of a single field with a
higher-derivative action. Let us start from the linearized Lagrangian
\beq\label{action-lin}
{\mathcal L}^{(2)}[h] = \frac{1}{2} h \Box h \, ,
\eeq
where $\Box= \nabla_\mu \nabla^\mu$ and our signature convention is $(-,+,+,+)$. One can
think of the field $h$ as that of the graviton with the factor of the inverse Newton's
constant $G^{-1}$ absorbed into the definition of $h$ to endow it with the canonical mass
dimension $[h]=1$. All tensor indices of $h$ have been discarded in order to make the
discussion as simple as possible.

Lagrangian (\ref{action-lin}) can be modified by introducing higher-derivative
non-renormalizable terms. For any finite number of such terms, the resulting theory has
additional propagating modes corresponding to the poles of the arising propagator.
However, when an infinite number of such higher-derivative terms is present, they
sometimes can be resummed into a denominator so that the only pole of the arising
propagator is still that of the massless graviton. One of the simplest examples realizing
this idea is
\beq\label{action-nl}
{\mathcal L} [h] = \frac{1}{2} h \left( \Box + \ell^2 \Box^2 + \ell^4 \Box^3 + \ell^6
\Box^4 + \ldots \right) h = \frac{1}{2} h \frac{\Box}{1 - \ell^2 \Box} h \, .
\eeq
Here, $\ell$ is the length scale of the modification. It is clear that the usual
``graviton'' is the only propagating mode in (\ref{action-nl}).  To see this explicitly,
note that the non-local theory (\ref{action-nl}) can be obtained from the local one
\beq\label{action-h-chi}
{\mathcal L} [h,\chi] = \frac{1}{2}\chi \left(\Box - m^2 \right) \chi +\chi \Box h
+\frac{1}{2} h\Box h \, ,
\eeq
where $m = \ell^{-1}$, by integrating out the auxiliary field $\chi$. Indeed, the
equation for $\chi$ that stems from (\ref{action-h-chi}) is formally solved by
\beq\label{chi-h}
\chi = \frac{\Box h}{m^2-\Box} \, .
\eeq
Substituting this into (\ref{action-h-chi}), one obtains (\ref{action-nl}). At the level
of action (\ref{action-h-chi}), the limit $m\to \infty$ that gives back
(\ref{action-lin}) works by making the field $\chi$ infinitely massive, and thus
effectively decoupling it from the theory. At the same time, writing action
(\ref{action-h-chi}) in terms of the new field variable $\tilde h = h + \chi$,
\beq \label{action-tilde}
{\mathcal L} [\tilde h,\chi] = \frac{1}{2} \tilde h \Box \tilde h - \frac{1}{2} m^2
\chi^2 \, ,
\eeq
one can see that the field $\chi$ is an auxiliary non-propagating field, and the theory
(\ref{action-h-chi}) has the same number of degrees of freedom and structure as the one
we started from, Eq.~(\ref{action-lin}). Thus, the described
modification (\ref{action-nl}) is just a field redefinition, which gives another way to
understand why no new propagating degrees of freedom is introduced in this
case.\footnote{One cannot discard the field $\chi$ in (\ref{action-tilde}) if the
``matter'' part of the action interacts with the original field $h \equiv \tilde h -
\chi$.  In this case, there arises a theory with non-propagating auxiliary scalar $\chi$
that couples to matter, resembling the modified source gravity \cite{Carroll:2006jn}.
This will also be the case in the modified-gravity theory under consideration in this
paper.}

The described elementary modification scheme is (almost) identical to the one at play in
the class of theories \cite{Krasnov:2006du} at the linearized level. However, in theories
\cite{Krasnov:2006du}, it is extended to a full non-linear level in a non-trivial
fashion. It then turns out that, by the same trick of introducing auxiliary
non-propagating fields, quite non-trivial terms in the arising effective metric
Lagrangian can be reproduced, for instance, the $Riemann^3$ term of importance in quantum
gravity at two loops; see \cite{Krasnov:2009ik}.

To summarize, the idea of the deformations of GR \cite{Krasnov:2006du} is that
higher-derivative terms in the effective
metric Lagrangian can be added without introducing extra degrees of freedom, and the
theory is then made local at the expense of introducing some non-propagating auxiliary
fields. In the class of theories \cite{Krasnov:2006du}, this idea is realized elegantly
by combining the metric and auxiliary fields into a single field --- an ${\rm SU}(2)$
Lie-algebra-valued two-form.

Another general comment should be made about the class of theories under investigation in
this paper. The above discussion seemingly implies that they can be regarded as
ultraviolet modifications of general relativity. Indeed, usually, modifications generated by the
addition of higher-derivative terms to the Hilbert--Einstein Lagrangian
are of relevance at high energies, e.g., close to the Planck scale,
but are unimportant at low energies of relevance in cosmology.  However, our above
example shows that this interpretation may not always be valid.
Indeed, the theory (\ref{action-nl}) has the field and dynamical content equivalent to
that of (\ref{action-lin}). This property will be lost after truncation of the series in
(\ref{action-nl}) or after the violation of the delicate relations between the
coefficients in this infinite series.  Through these relations, the high-derivative terms
(potentially important in the ultraviolet) are all connected with the low-derivative
terms (important in the infrared).  In the case of the modified gravity theory to be
considered, this UV/IR interplay will be seen in the fact that the theory can be
generated from the self-dual formulation of general relativity simply by making the
cosmological constant a function of the variable which in GR has the meaning of the Weyl
curvature (more on this below). This allows gravity to be modified on cosmological scales
rather naturally, simply by making the usual cosmological constant a slowly varying
function of the curvature. At the same time, it is easy to remain consistent with GR on
Solar-System scales by requiring the ``cosmological function'' to be approximately
constant in the relevant region of curvatures (see the next section for more details). In
this sense, the theory simultaneously looks like an infrared modification of gravity.
This interplay between the ultraviolet and infrared modification is an interesting
feature that characterizes the class of theories under investigation.

The plan of the paper is as follows. In Sec.~\ref{sec:family}, we briefly describe the
family of deformations of GR which is the subject of this study. In Sec.~\ref{sec:hom},
we describe the homogeneous isotropic Universe and the usual Friedmann equations in the
language of two-forms adopted in this paper. Section \ref{sec:lin} describes the
linearized field equations around the evolving cosmological background. In
Sec.~\ref{sec:class}, we classify the types of perturbations and describe the physical
gauge-invariant quantities. Section \ref{sec:scalar} obtains equations describing
perturbations of the scalar type. These are analyzed in details in
Sec.~\ref{sec:evolution}. We study the effect of modification on the CMB and matter power
spectrum in Sec.~\ref{sec:power}. We conclude with a brief discussion of the results
obtained. Preliminary results concerning tensor modes are described in
Appendix~\ref{sec:tens}.

\section{A family of deformations of GR}
\label{sec:family}

In this section, we briefly review the theory under investigation in this work. For more
details, the reader is directed to a recent description in \cite{Krasnov:2009ik}.

\subsection{Preliminaries: two-form field and the metric}\label{prelimin}

As we mentioned in the introduction, in our theory the metric and the auxiliary
non-propagating fields are ``unified'' into a single two-form field $B^i_{\mu\nu}$,
$i=1,2,3$, where $\mu,\nu$ are spacetime indices. The (complexified) rotation group ${\rm
SO}(3,\C)$ acts on the two-form $B^i_{\mu\nu}$ via $B^i_{\mu\nu}\to O^i_j B^j_{\mu\nu}$,
$O^i_j\in {\rm SO}(3,\C)$, and the theory is invariant under the corresponding gauge
transformations as well as under the action of spacetime diffeomorphisms.

The physical metric of the spacetime, to which all fields are supposed to couple
universally, is determined uniquely by the two-form field. To begin with, we require that
the triple of two-forms $B^i_{\mu\nu}$ be self-dual with respect to this metric, which
determines its conformal class. In other words, introducing the associated volume form
$\epsilon_{\mu\nu\rho\sigma}$ for a metric $g_{\mu\nu}$ and imposing the self-duality
conditions\footnote{The spacetime indices in (\ref{sdual}) are raised with the use of the
inverse metric $g^{\mu\nu}$.}
\beq \label{sdual}
\frac{1}{2} \epsilon_{\mu\nu}^{\quad \rho\sigma} B^i_{\rho\sigma} = \im B^i_{\mu\nu}
\eeq
fixes the metric $g_{\mu\nu}$ modulo a conformal rescaling.

An equivalent explicit description of the conformal class of metrics determined by
$B^i_{\mu\nu}$ is given by the formula
\beq\label{Urb}
\sqrt{-{\rm det}\left( g_{\mu\nu} \right)} \, g_{\mu\nu} \propto
\tilde{\epsilon}^{\alpha\beta\gamma\delta} B^i_{\mu\alpha} B^j_{\nu\beta}
B^k_{\gamma\delta} \epsilon^{ijk} \, ,
\eeq
where $\tilde \epsilon^{\alpha\beta\gamma\delta}$
is a densitized completely anti-symmetric tensor having components $\pm 1$ in any
coordinate system, and the proportionality symbol is used here to denote that the metric
is defined up to conformal rescalings. The reality conditions imposed on the theory are
that the conformal metric (\ref{Urb}) is real Lorentzian. To fix a unique metric from its
conformal class, it suffices to specify its volume form. Below, we shall explain how the
triple $B^i_{\mu\nu}$ defines not only a conformal class of metrics, but a volume form as
well.

Vice versa, given a real metric $g_{\mu\nu}$ with Lorentzian signature, one can introduce
the corresponding tetrad one-forms $ \theta^I_\mu$, $I=0,1,2,3$, so that the line element
reads
\beq
ds^2 = \theta^I \otimes \theta^J \eta_{IJ}\, ,
\eeq
where $\eta_{IJ}$ is the Minkowski metric. Then, introducing an arbitrary time plus space split
$I=(0,a)$ of the "internal" index $I$, we can define a set of self-dual metric two-forms
$\Sigma^{a}\,$:
\beq
\Sigma^{a} = \im \theta^0 \wedge \theta^{a} - \frac{1}{2} \epsilon^{abc} \theta^{b}\wedge
\theta^{c}\, .
\eeq
Here, the letters $a,b,c$ are new ${\rm SO}(3)$ indices. It is convenient to distinguish
between the original ${\rm SO}(3)$ indices in $B^i$ and the new index that appears in the
metric two-forms $\Sigma^{a}$, which justifies the notation. Let us also give an
expression for a convenient basis in the space of anti-self-dual two-forms:
\beq\label{Sigma-bar}
\bar{\Sigma}^{a} = \im \theta^0 \wedge \theta^{a} + \frac{1}{2} \epsilon^{abc}
\theta^{b}\wedge \theta^{c} \, .
\eeq

Any other set of self-dual two-forms can be decomposed in the basis provided by
$\Sigma^{a}$. Thus, we can write
\beq\label{B-Sigm}
B^i = b^{i}_a \Sigma^{a}\, .
\eeq
Therefore, the main dynamical field of the theory can always be chosen as a collection of
a metric $g_{\mu\nu}$ that, in turn, defines the metric forms $\Sigma^{a}$, together with
the nine scalars forming the matrix $b^{i}_a$. These scalars will later be seen to
be non-dynamical and determined by other fields present in the system. We
do not impose any independent reality conditions on $b^i_a$ as such reality conditions are
induced once $b^i_a$ are determined in terms of other fields. Thus, $b^i_a$
are in general complex fields. Note that, as the metric $g_{\mu\nu}$
undergoes a conformal transformation $g_{\mu\nu}\to \Omega^2 g_{\mu\nu}$, the metric
self-dual two-forms transform as $\Sigma^{a}\to \Omega^2 \Sigma^{a}$. Thus, if the
scalars $b^{i}_a$ transform as $b^{i}_a\to \Omega^{-2} b^{i}_a$, then the original
two-forms $B^i$ are unchanged. A certain normalization condition on the scalars $b^{i}_a$
will be introduced below to fix this conformal freedom, thus fixing a particular metric
from the conformal class of (\ref{Urb}).

\subsection{The gravitational action}

The vacuum theory of gravity under consideration is described by the following action:
\beq\label{action}
S[B,A]=\frac{\im}{8\pi G} \int \left[ B^i\wedge F^i(A) - \frac{1}{2} V \left( B^i\wedge
B^j \right) \right] \, .
\eeq
Here, $A^i$ is an ${\rm SO}(3,\C)$ connection field,
$F^i(A)=dA^i+(1/2)\epsilon^{ijk}A^j\wedge A^k$ is its curvature, and $V(M^{ij})$ is a
(holomorphic) function of a complex symmetric $3 \times 3$ matrix variable $M^{ij}$. It
is required to be a homogeneous function of degree one: $V\left(\alpha
M^{ij}\right)=\alpha V\left( M^{ij} \right)$ for any number $\alpha$, so that it can be
applied to a matrix-valued $B^i\wedge B^j$ four-form, with the result being again a
four-form; see \cite{Krasnov:2009ik} for more details. The function $V(M^{ij})$ is also
required to be a scalar with respect to the ${\rm SO}(3,\C)$ gauge transformations. In
action (\ref{action}), this function plays the role of potential for the $B^i$ field.

Using representation (\ref{B-Sigm}), let us introduce the ``internal metric''
\beq \label{int-met}
h^{ij}:= b^{i}_a b^{j}_b \delta^{ab}\, ,
\eeq
which is a symmetric $3\times 3$ matrix, in general complex. It turns out to be useful to
parameterize it by its trace $h = h^{ij} \delta_{ij}$ and trace-free part $H^{ij}\,$:
\beq
h^{ij} = \frac13 h \left( \delta^{ij} + H^{ij} \right)\, .
\eeq
Then the potential $V \left( B^i\wedge B^j \right)$ gives rise to the ${\rm
SO}(3,\C)$-invariant function $V \left( h^{ij} \right)$, which can be presented as
follows:
\beq\label{V}
V \left( h^{ij} \right) = \frac13 h\, U(H), \qquad U(H)=\Lambda_0 - \frac{1}{8 \ell^2}
{\rm Tr} H^2 + {\cal O} \left( H^3 \right) \, .
\eeq
Here, $U(H)$ is a function of the traceless matrix $H^{ij}$ with dimensions of curvature,
and we have expanded it in powers of $H^{ij}$ assuming that this potential function is
analytic in the neighborhood of $H^{ij}=0$. Below we will see that the case of $h^{ij} =
\delta^{ij}$ corresponds to general relativity, so that the traceless matrix $H^{ij}$
parameterizes the deformation away from GR. The constant term $\Lambda_0$ in
decomposition (\ref{V}) will play the role of the cosmological constant. The quantity
$\ell$ is a new parameter of the theory with dimensions of length. In principle, in the
theory under consideration, there is scope for making this parameter complex, but, in
this paper, we shall only consider the simplest case of real $\ell$. The numerical
constant in the expansion in (\ref{V}) is introduced for future convenience. We will see
that only negative sign of the second term in (\ref{V}) describes a theory without
instabilities, which explains our choice of this sign.

\subsection{Coupling to matter}

As we have stated in Sec.~\ref{prelimin}, the physical metric which universally couples
to matter belongs to the conformal class determined by the two-form field $B^i$.  It is
necessary to fix the remaining conformal freedom and to specify this metric uniquely.

Under conformal transformations, described in the end of Sec.~\ref{prelimin}, the
``internal metric'' introduced in (\ref{int-met}) transforms as $h^{ij}\to \Omega^{-4}
h^{ij}$. One can specify the physical metric by fixing the conformal ambiguity for
$h^{ij}$. To do this, we introduce another scalar (holomorphic) function $R\left( h^{ij}
\right)$ of the matrix $h^{ij}$ that is required to be homogeneous of degree one in the
components of $h^{ij}$: $R \left(\alpha h^{ij} \right) = \alpha R \left( h^{ij} \right)$.
The conformal freedom of $h^{ij}$ is then fixed by the condition
\beq\label{conf-fix}
R \left( h^{ij} \right) = 1\, .
\eeq
While this prescription for fixing the physical metric may seem arbitrary at first, it
can be shown to arise quite naturally by considering the motion of a ``small body'' in
the theory described by (\ref{action}); see \cite{Krasnov:2008ui}.

We normalize the function $R \left( h^{ij} \right)$ so that $R \left( \delta^{ij} \right)
= 1$.  The potential $R \left( h^{ij} \right)$ can then be decomposed similarly to
(\ref{V}):
\beq\label{R}
R(h) = \frac13 h\, U_m (H)\, , \qquad U_m(H) = 1 - \frac{g}{2} {\rm Tr} H^2 + {\cal O}
\left( H^3 \right) \, ,
\eeq
where $U_m(H)$ is a new function of the traceless matrix $H \equiv H^{ij}$, again assumed
to be analytic in the neighborhood of $H^{ij}=0$. The quantity $g$ is a dimensionless
parameter. Just as with the case of $\ell$, in the theory under consideration, there is
scope for making the parameter $g$ complex. However, in this paper, we shall only
consider the simplest case of real $\ell$ and $g$. Note that, once the condition
(\ref{conf-fix}) is imposed, one can use it to express the trace part of the matrix
$h^{ij}$ in terms of its traceless part.

We shall see that only the terms written in (\ref{V}) and (\ref{R}) are going to matter
for the linearized theory we are going to consider. Thus, the only new parameters that we
will have to consider on top of those available in GR are the length parameter $\ell$ in
(\ref{V}) and the dimensionless constant $g$ in (\ref{R}) that can be of any sign.

In the class of theories under consideration, matter fields will in general couple to the
two-form $B^i$ that plays the role of the fundamental geometrical field of the theory.
However, for the purposes of this paper, we assume that our matter content is
``standard'' in the sense that our material components couple to the spacetime metric in
the conformal class defined by $B^i$ and further selected by the condition
(\ref{conf-fix}). Thus, we assume that the dependence of the matter action $S_m$ on $B^i$
is such that $S_m[B]$ is a functional of $g_{\mu\nu}$ and depends on the scalars $b^i_a$
solely via the function $R(h)$. This allows us to write:
\beq
\delta S_m = \int \sqrt{-{\rm det} \left( g_{\mu\nu} \right)} \left( T_{\mu\nu} \delta
g^{\mu\nu} + \frac{\partial S_m}{\partial R} \frac{\partial R}{\partial b^i_a} \delta
b^i_a \right).
\eeq
However, since the two-form field $B^i$ does not transform under the conformal
transformations $\delta g_{\mu\nu} = \varepsilon g_{\mu\nu}$, $\delta b^i_a = -
\varepsilon b^i_a$, any material action that arises from $S_m[B]$ is also
conformally-invariant. This immediately gives:
\beq
T = \frac{\partial S_m}{\partial R} \frac{\partial R}{\partial b^i_a} b^i_a,
\eeq
where $T:=T_{\mu\nu} g^{\mu\nu}$, or, using the homogeneity of $R(h)$:
\beq
\frac{\partial S_m}{\partial R}=\frac{T}{2R}.
\eeq
In other words, the variation of the material action contains the following two terms:
\beq
\delta S_m = \int \sqrt{-{\rm det}\left( g_{\mu\nu} \right)} \left( T_{\mu\nu} \delta
g^{\mu\nu} + \frac{T}{2R} \frac{\partial R}{\partial b^i_a} \delta b^i_a \right).
\eeq
After the condition (\ref{conf-fix}) is imposed, the quantity $T_{\mu\nu}$ becomes the
usual stress-energy tensor of matter. We can anticipate already at this stage that the
field equations will contain terms involving the first derivative $\partial R/\partial
b^i_a$ of the potential $R(h)$.

\subsection{Field equations}

The set of equations obtained by varying the full action of the theory (given by
(\ref{action}) plus the material action) with respect to $A^i$ (assuming that the
material action does not depend on $A^i$) is
\beq\label{compat}
dB^i + \epsilon^{ijk} A^j(B)\wedge B^k=0 \, .
\eeq
These equations are solved by
\beq\label{A-forms}
A^i_{\mu}(B) := \frac{1}{2\, {\rm det\,} B} B^{i\rho\sigma} B^j_{\rho\mu} \nabla^\nu
B^j_{\nu\sigma}\, , \qquad {\rm det\,} B :=-\frac{1}{24} \epsilon^{ijk} B^i_\mu{}^\nu
B^j_\nu{}^\rho B^k_\rho{}^\mu\, .
\eeq
We note that this connection is conformally-invariant and has the correct transformation
properties of an ${\rm SO}(3)$ connection. That is, when the two-form field
$B^i_{\mu\nu}$ transforms as $\delta B^i_{\mu\nu}=\epsilon^{ijk} \omega^j B^k_{\mu\nu}$,
the connection transforms as $\delta_\omega A^i_\mu = \epsilon^{ijk}\omega^j A^k_\mu -
\partial_\mu \omega^i$. A demonstration of this involves a
simple identity satisfied by the two-forms $B^i_{\mu\nu}$; see \cite{Krasnov:2009ik}. We
also note that, in most cases, instead of using (\ref{A-forms}) for finding $A^i(B)$, it
is easier to solve (\ref{compat}) directly.

Once the connection $A^i(B)$ is found, one computes its curvature:
\beq
F^i:= dA^i + \frac{1}{2} \epsilon^{ijk} A^j\wedge A^k\, .
\eeq
The field equations then read, in the parametrization (\ref{B-Sigm}):
\beq\label{feqs}
b^{i}_a F^i\left[A (b\Sigma) \right] = \left( \frac{\partial U}{\partial H^{ij}} b^{i}_a
b^{j}_b + \frac{1}{3} \Lambda b^i_a b^i_b \right) \Sigma^{b} +  2\pi G T \left(
\frac{\partial U_m}{\partial H^{ij}} b^{i}_a b^{j}_b + \frac{1}{3} \Lambda_m b^i_a b^i_b
\right) \Sigma^{b} - 2\pi G T_{ab} \bar \Sigma^{b} \, .
\eeq
Here, $\Lambda$ and $\Lambda_m$ are the Legendre transforms
\beq\label{Legendre}
\Lambda = U(H)- \frac{\partial U}{\partial H^{ij}} H^{ij}\, , \qquad \Lambda_m = U_m(H)-
\frac{\partial U_m}{\partial H^{ij}} H^{ij}
\eeq
of the potentials $U(H)$ and $U_m(H)$, respectively, $T$ is the trace of the
stress-energy tensor of matter, $T_{ab}$ are the components of the traceless part of the
stress-energy tensor, and $\bar{\Sigma}^{a}$ are the anti-self-dual metric two-forms
(\ref{Sigma-bar}). For the matter in the form of ideal fluid, we have
\beq
T = (\rho - 3 p)\, , \qquad T_{ab} = (\rho + p) \left( \frac{\delta_{ab}}{1-|u|^2} + 2
\im \epsilon_{abc} \frac{u^c}{\sqrt{1-|u|^2}}\right) - 3 \cS_{ab}\, ,
\eeq
where $\rho$ and $p$ are the energy density and pressure, respectively, $u^a$ is
(proportional to) the momentum vector, and $\cS_{ab}$ is the traceless matrix describing
shear.

As anticipated in the previous subsection, the right-hand side of our field equations
includes a term proportional to the trace $T$ of the stress-energy tensor and containing
the first derivative of the potential function $R(h)$ (or $U_m(H)$) of the material
sector. The remainder of the dependence of the right-hand side of (\ref{feqs}) on the
stress-energy tensor components can be easily understood by working out how matter
couples to gravity in the so-called Pleba\'nski formulation of GR; see, e.g.,
\cite{Krasnov:2009pu}. The only change in the present case as compared to GR in the
Pleba\'nski formulation is in the appearance of the derivative terms $\partial
U_m(H)/\partial H$. A detailed derivation of the field equations (\ref{feqs}) can be
found in \cite{Krasnov:2008ui}.

In (\ref{Legendre}), we have introduced the ``cosmological function'' $\Lambda$ as the
Legendre transform of $U(H)$. When $H^{ij}=0$, the above field equations reduce to those
of GR in Pleba\'nski formulation. For
later purposes, we note that the function $\Lambda$ is, in fact, a function $\Lambda
(\Psi)$ of the quantity
\beq
\Psi \equiv \Psi^{ij}:=\frac{\partial U}{\partial H^{ij}}
\eeq
that can be related to the usual Weyl curvature (at least for small deviations from GR).
It is equally possible to parameterize the theory either by the original potential $U(H)$
or by its Legendre transform $\Lambda(\Psi)$. However, as we shall discuss in more detail
a little below, the nature of the modification becomes more clear in the $\Lambda(\Psi)$
parametrization. This is due to the fact that the field $H^{ij}$ that measures the
departure from GR arises in this parametrization as the first derivative
\beq\label{H-Lambd}
H^{ij} = - \frac{\partial \Lambda}{\partial \Psi^{ij}}
\eeq
of the cosmological function $\Lambda$. Thus, for a function $\Lambda (\Psi)$ which is
constant in its domain to a high degree, the deviations from GR are small.

\subsection{Large-distance modifications}

As we have already mentioned in the introduction, the class of theories under
investigation can be viewed as ultraviolet modifications of gravity. This is most clearly
seen in the parametrization of the theory by the potential $U(H)$. When it is expanded
around $H^{ij}=0$, the first non-trivial term contains a mass parameter $m^2=1/\ell^2$,
and modifications away from GR only manifest themselves at energies $E$ larger than $m$.
The low-energy limit $E\ll m$ of any of our theories is given by GR\@. However, when the
theory is looked at non-perturbatively, e.g., in the Hamiltonian formulation (see
\cite{Krasnov:2007cq}), one finds that the parametrization by the Legendre transform
$\Lambda(\Psi)$ of $U(H)$ is more natural. More specifically, one finds that the
departure of the theory from GR is connected with the fact that the cosmological constant
becomes a function $\Lambda(\Psi)$ of what in GR used to be the Weyl part of the Riemann
curvature tensor. The parametrization by $\Lambda(\Psi)$ then allows one to arrange for
modification of gravity at any desirable scale of curvatures while keeping it close to GR
at the scales of the Solar System. Indeed, the theory behaves close to GR whenever the
first derivative (\ref{H-Lambd}) of the function $\Lambda(\Psi)$ is small. Thus, to be
consistent with the Solar-System tests of gravity, one should assume the function
$\Lambda(\Psi)$ to be approximately constant in the range of curvatures probed in the
solar neighborhood. One can then allow $\Lambda(\Psi)$ to vary in the range of curvatures
much smaller than those in the Solar System, which will produce deviations from GR at
large distances (see \cite{Krasnov:2007ky}  as well as a more accessible exposition \cite{Krasnov:2008sb}).
Thus, according to our modification
scenario, the value of the cosmological constant in the Solar System can be different
from the value $\Lambda_0$ measured in cosmology, with the function $\Lambda(\Psi)$
interpolating between these two values. The parameter $\ell^2$ arising in (\ref{V}) is
then given by the second derivative of the function $\Lambda(\Psi)$ at $\Psi^{ij}=0$.
Since gravity is not tested directly at very large distances, there are only very weak
direct constraints on this parameter. As we shall see, a study of the physics of the
cosmic microwave background (CMB) and formation of the large-scale structure as predicted
by this theory will place much stronger constraints on the value of $\ell$.

Another important point that we would like to make is in the following. A detailed
analysis shows that, as the first derivative (\ref{H-Lambd}) of $\Lambda(\Psi)$ becomes
of order unity (in the sense that any of the eigenvalues of the symmetric traceless
matrix $H^{ij}$ become of order one), the theory under consideration develops a specific
degeneracy, which results in a singularity of metric (\ref{Urb}) \cite{Krasnov:2007ky}.
It can also be argued that it is no longer consistent to use the classical theory in this
regime because quantum effects become important. Thus, it is necessary to avoid this
regime if one wants to remain in the domain of applicability of the classical physics,
which is usually the case for large distances of relevance for cosmology. Therefore, the
classical theory will remain applicable only when the condition
\beq\label{strong-coupl}
\left| \frac{\partial \Lambda}{\partial \Psi^{ij}} \right| \ll 1
\eeq
is satisfied. This condition can be rephrased by saying that the departure from GR as
measured by the quantity (\ref{H-Lambd}) must be small. We assume this condition in what
follows.

\subsection{The cosmological constant and vacuum energy}\label{sec:lambda}

In general relativity, the cosmological constant $\Lambda$ can be regarded as
corresponding to the energy density of the vacuum, and one is free to absorb it in the
energy density and pressure or into the scalar-field potential $V (\varphi)$ in all
equations without exception. That is, the transformation
\beq \label{shift}
\Lambda \to \Lambda - \lambda \, , \qquad \rho \to \rho + \frac{\lambda}{8 \pi G} \, ,
\qquad p \to p - \frac{\lambda}{8 \pi G}
\eeq
in the case of ideal fluid or
\beq \label{shiftin}
\Lambda \to \Lambda - \lambda \, , \qquad V (\varphi) + \frac{\lambda}{8 \pi G}
\eeq
in the case of scalar field leaves invariant all equations of the theory.

In the theory under consideration, this property is true only for the equations
describing the homogeneous and isotropic universe, which are the same as in general
relativity.  It is not true in the general case, which can be seen in (\ref{feqs}), in
which the trace of the stress-energy tensor of matter $T$ enters differently from the
genuine cosmological constant $\Lambda_0$ [present in the ``cosmological function''
$\Lambda(\Psi)$].  This feature of the theory of modified gravity persists in the
equations for cosmological perturbations, as we shall see below. An exception is the
special case when the parameter $\ell^2$ is fixed to be $\ell^2 = 1/4 g \Lambda_0$.
In this case, the invariance with respect to (\ref{shift}) is restored at the level of linear
perturbations.

\section{Homogeneous and isotropic universe}
\label{sec:hom}

Consider a homogeneous and isotropic background described by the two-forms $B_0^i$. On
such a background, the traceless matrix $H^{ij}$ describing the departure from GR is
equal to zero due to the spacetime symmetries. Thus, we have $H^{ij}_0 = 0$, and we can
choose the scalars $b^{i}_{0a}=\delta^{i}_a$, so that $B^i_0=\Sigma_0^i$. The background
equations are then the Friedmann equations in the Pleba\'nski formalism:
\beq\label{backgr-feqs}
F^i \left[ A(\Sigma_0) \right] = \frac{2\pi G T_0 + \Lambda_0}{3} \Sigma^i_0  - 2\pi G
\bar{T}_0 \bar{\Sigma}^j_0 \, .
\eeq
Here,
\beq
T_0 := (\rho - 3p), \qquad \bar{T}_0 := (\rho + p)
\eeq
are the components of the background stress-energy tensor. We emphasize that the
background evolves in our theory in exactly the same way as in GR\@. Modifications only
arise in the dynamics of perturbations. This is, however, the story for an ideally
homogeneous backgrounds. In reality, there are inhomogeneities on various scales that
need to be averaged over. It is natural to expect that this will introduce modifications
of the background equations as well, to be taken into account. However, since the
averaging issue is not fully understood even in the simpler case of GR, we postpone such
an analysis to our future publications, and, in this paper, study the cosmological
perturbations around an ideally homogeneous and isotropic background.

In the conformal time $\eta$, the metric of a spatially flat homogenous and isotropic
universe is:
\beq\label{metric}
ds_0^2 = a^2(\eta)\left[ d\eta^2 - \sum_i \left(dx^i\right)^2 \right] \, .
\eeq
In what follows, we often omit the argument of the scale factor $a(\eta)$.

A set of self-dual and anti-self-dual two-forms describing this metric can be chosen as
\beq\label{B-background}
\Sigma^i_0 = a^2 \left( \im d\eta\wedge dx^i - \frac{1}{2} \epsilon^{ijk} dx^j \wedge
dx^k \right)\, , \quad \bar{\Sigma}^i_0 = a^2 \left( \im d\eta\wedge dx^i + \frac{1}{2}
\epsilon^{ijk} dx^j \wedge dx^k \right)\, .
\eeq
The connection $A^i_0$ compatible with the set of self-dual two-forms $\Sigma^i_0$, i.e.,
satisfying $d\Sigma^i_0 + \epsilon^{ijk}A^j_0 \wedge \Sigma^k_0=0$, is given by
\beq\label{A-background}
A^i_0 = \im \cH dx^i \, ,
\eeq
where
\beq
\cH := \frac{a'}{a} \, ,
\eeq
and the prime denotes the derivative with respect to the conformal time $\eta$. The
curvature $F^i = dA^i + \frac12 \epsilon^{ijk} A^j \wedge A^k$ of connection
(\ref{A-background}) is given by
\beq\label{F-background}
F_0^i = \frac{1}{2a^2} \left( \cH' + \cH^2 \right) \Sigma^i + \frac{1}{2a^2} \left( \cH'
- \cH^2 \right) \bar{\Sigma}^i \, ,
\eeq
where we have used the relations
\beq\label{two-form-B}
d\eta\wedge dx^i = \frac{1}{2 \im a^2} \left( \Sigma^i_0 + \bar{\Sigma}^i_0 \right)\, ,
\qquad dx^i\wedge dx^j = \frac{1}{2a^2} \epsilon^{ijk} \left( \bar{\Sigma}^k_0 -
\Sigma^k_0 \right)
\eeq
to decompose $F^i_0$ into its self-dual and anti-self-dual parts.

Substituting (\ref{F-background}) into (\ref{backgr-feqs}), we obtain the dynamical
equations for the background. The condition that the self-dual and anti-self-dual parts
of the field equations are separately satisfied gives two equations whose symbolic linear
combinations $\Sigma-\bar{\Sigma}$ and $3\Sigma + \bar{\Sigma}$ are just the Friedmann
equations
\beq\label{Friedmann}
\frac{\cH^2}{a^2} = \frac{8\pi G \rho}{3} + \frac{\Lambda_0}{3} \, , \qquad \frac{2\cH' +
\cH^2}{a^2} = -8\pi G p + \Lambda_0 \, .
\eeq

\section{Linearized field equations}
\label{sec:lin}

Linearizing the field equations (\ref{feqs}) around the homogeneous and isotropic
background, we denote the perturbation of a quantity by a symbol $\delta$ next to the
symbol representing it. There arises a perturbation of the metric two-forms $\delta
\Sigma^{i}$ as well as a perturbation of the scalars $\delta b^{i}_a$. Using the fact
that $B^{i}_{0a}=\delta^{i}_a$, now we can identify the two types of SO(3) indices. Thus,
from now on we drop the distinction between the two types, and use only the $i,j,k$
indices. The condition $R\left( h^{ij} \right) = 1$ implies, at the first order in
perturbation, that the matrix $\delta b^{ij}$ is traceless.  The perturbations of the
metric $h^{ij}$ and of its trace-free part are then $\delta h^{ij} = \delta H^{ij} = 2
\delta b^{(ij)}$. Since $H^{ij}_0 = 0$, the first-order perturbation of the quantity
${\rm Tr} H^2$ that appears in the potential functions is zero. Thus, $\delta \Lambda =
0$ and $\delta \Lambda_m = 0$.

After some algebra, the linearized field equations are found to be
\beq\label{lin-feqs}
\delta F^i + \delta b^{ij} F^j_0 = \frac{2\pi G}{3} \delta T \Sigma^i_0 - 2\pi G \delta
T^{ij} \bar \Sigma^j_0 + \frac{2\pi G T_0 + \Lambda_0}{3} \delta \Sigma^i - 2\pi G
\bar{T}_0 \delta \bar{\Sigma}^i - 4\pi G \kappa  \, \delta b^{ij} \Sigma^j_0\, .
\eeq
Only the last terms on the left-hand side and on the right-hand side of this equation are
not present in GR, with all other terms being exactly like in the Pleba\'nski formulation
of GR\@. The perturbation of the components of the stress-energy tensor of fluid is given
by
\beq\label{delta-T}
\delta T = \delta \rho - 3 \delta p\, , \qquad \delta T^{ij} = (\delta\rho + \delta p)
\delta^{ij} + 2 \im \epsilon^{ijk} \delta u^k - 3 \delta \cS^{ij}\, ,
\eeq
where, for convenience, we have absorbed the factor $\rho + p$ into the momentum vector
$u^i$. The quantity $\delta F^i$ is the linearized curvature corresponding to the
perturbation of the $B^i$ field given by
\beq
\delta B^i = \delta b^{ij} \Sigma^j_0 + \delta \Sigma^i\, ,
\eeq
where $\delta \Sigma^i$ is the perturbation of the metric self-dual two-forms (see below
for explicit expressions). In (\ref{lin-feqs}), we have introduced a new parameter
$\kappa$ with dimensions of the energy density:
\beq \label{kappa}
\kappa := \left( g - \frac13 \right) T_0 + \frac{1}{8 \pi G } \left( \frac{1}{\ell^2} -
\frac43 \Lambda_0 \right) = \beta ( \rho - 3 p) + \gamma \rho_\Lambda \, ,
\eeq
where $T_0=\rho - 3 p$ is the trace of the material stress-energy tensor, $\rho_\Lambda =
\Lambda_0/ 8\pi G$ is the energy density of the cosmological constant, and $\ell$ and $g$
are defined in (\ref{V}) and (\ref{R}), respectively. The quantities $\beta$ and $\gamma$
are two new convenient dimensionless parameters:
\beq \label{bg}
\beta = g- \frac{1}{3}, \qquad \gamma = \frac{1}{\ell^2 \Lambda_0} - \frac43 \, .
\eeq
These parameters have the following meaning. The parameter $\gamma$ determines the
deviation of the gravitational potential $V \left( h^{ij} \right)$ from one
given by $V \left( h^{ij} \right) =\Lambda_0 R \left( h^{ij}\right)$. This case corresponds to
$4\beta=\gamma$ and describes the theory that couples in the same way to the
cosmological constant and to the trace of the stress-energy tensor. The parameter $\beta$
controls the deviation of the physical metric from the so-called Urbantke metric which
coincides with (\ref{Urb}) precisely. The value $\beta = 0$ corresponds to coupling of
matter to the Urbantke metric. Finally, the limit to GR is obtained by keeping $\beta$
fixed (and arbitrary), and sending $\ell \to 0$ (thus $\gamma \to \infty$) keeping the
product $\gamma \,\delta B^{ij}$ fixed. To summarize, schematically we have:
\beq
\beta= 0 \ \Longleftrightarrow \ \mbox{coupling to the Urbantke metric}\, , \qquad
\gamma\to\infty \ \Longleftrightarrow \ \mbox{limit to GR} \, .
\eeq

Now, using the expression (\ref{F-background}) for the background curvature and equations
(\ref{backgr-feqs}), we get
\beq
\frac{2\pi G T_0+\Lambda_0}{3} = \frac{\cH^2 + \cH'}{2a^2} \, , \qquad 2\pi G\bar{T}_0 =
\frac{\cH^2 - \cH'}{2a^2} \, .
\eeq
The linear equations for perturbations can then be written in the form
\ber
&& \delta B^{ij} F^j_0 + \delta F^i - \frac{1}{2a^2}(\cH^2+\cH') \delta \Sigma^i +
\frac{1}{2a^2}(\cH^2-\cH')\delta \bar{\Sigma}^i \nonumber \\ && = \frac{2\pi G}{3} \delta
T \Sigma^i_0 - 2\pi G \delta T^{ij} \bar{\Sigma}^j_0 - 4\pi G\kappa\, \delta B^{ij}
\Sigma^j_0 \, , \label{lin-feqs*}
\eer
where the stress-energy perturbations of the fluid are given, as before, by
\beq\label{delta-T*}
\delta T = (\delta \rho-3\delta p), \qquad \delta T^{ij} = (\delta\rho + \delta p)
\delta^{ij} + 2 \im \epsilon^{ijk}\delta u^k - 3 \delta \cS^{ij}\, ,
\eeq
and $\delta B^{ij}$ is the non-metric part of the perturbation (see the next section for
explicit expressions). This form of the linearized equations is most convenient for
further calculations.

\section{Classification of perturbations}
\label{sec:class}

A general infinitesimal perturbation of the background (\ref{B-background}) can be
decomposed into the background two-forms:
\beq\label{perturb-general}
\delta B^i = \delta \Phi^{ij} \Sigma^j_0 + \delta \Upsilon^{ij} \bar{\Sigma}^j_0 \, .
\eeq
However, it is clear that much of this perturbation is a gauge transformation. Indeed,
the anti-symmetric part $\delta\Phi^{[ij]}$ describes an infinitesimal ${\rm SO}(3)$
rotation of the triple of two-forms $\Sigma^i$ and is thus a pure gauge. Moreover, some
part of the general perturbation (\ref{perturb-general}) is an infinitesimal
diffeomorphism of the background (\ref{B-background}). To separate the physical
quantities from the gauge ones, we thus need to compute the effect of an infinitesimal
diffeomorphism on the background (\ref{B-background}). The Lie derivative of $B^i$,
suitably corrected by a gauge transformation, can be shown to be given by $\cL_\xi B^i =
\cD B^i (\xi)$, where $B^i (\xi) $ is the interior multiplication of a vector field $\xi$
with two-form $B^i$, and $\cD$ is the covariant derivative with respect to the
$B$-compatible connection. For the background (\ref{B-background}), a straightforward but
lengthy calculation gives
\ber \label{diffeo}
\cL_\xi \Sigma^i_0 &=& \frac{1}{2} \left[ 4 \cH \xi^0 + \xi^0{}' + \Delta \zeta \right]
\Sigma_0^i + \xi_{[i,j]} \Sigma_0^j + \frac{\im}{2} \epsilon^{ijk} \left[ 2\cH (\xi_j +
\zeta_{,j}) +
(\xi_j+\zeta_{,j})' + \xi^0{}_{,j}\right] \Sigma^k_0 \nonumber \\
&& {} + \frac{1}{2} \left[ \xi^0{}' - \Delta \zeta \right] \bar{\Sigma}_0^i + \left[
\xi_{(i,j)} + \zeta_{,ij} \right] \bar{\Sigma}_0^j + \frac{\im}{2} \epsilon^{ijk} \left[
(\xi_j+\zeta_{,j})' - \xi^0{}_{,j}\right] \bar{\Sigma}^k_0 \, .
\eer
Here, we have decomposed the vector field $\xi$ into its components $\xi= \left(\xi^0,\,
\xi_i+\zeta_{,i}\right)$, where $\xi_i$ is transverse $\xi^i{}_{,i}=0$. We see that the
effect of a diffeomorphism is a change in the metric described by $\Sigma^i_0$ together
with a gauge transformation on $\Sigma^i_0$. The latter can always be corrected by an
${\rm SO}(3)$ rotation, which allows us to drop the anti-symmetric part of the tensor
$\delta\Phi^{ij}$ from now on. Note that diffeomorphisms do not affect the symmetric
traceless part of $\delta\Phi^{ij}$, i.e., they do not act on the non-metric\footnote{We
shall often refer to the non-dynamical scalars present in $B^i$ in addition to the metric
as its ``non-metric'' components.} components of the $B^i$ field. In principle, this
could have been expected, for they act in the spacetime manifold, not in the internal
space, where the non-metric components of $B^i$ reside.

Let us now consider the usual classification of perturbations into scalar, vector and
tensor sectors and deduce the most general form of these perturbations after all gauge
freedom is fixed.

\subsection{Scalar sector}

Scalar perturbations are described by scalar functions, so that we have
\ber
\delta \Phi^{ij} = \phi \delta^{ij} + \frac{1}{a^2} \left( \chi_{,ij} - \frac{1}{3}
\delta_{ij} \Delta \chi \right) + \frac{\im}{4}\epsilon^{ijk} \theta_{,k}\, , \\
\delta \Upsilon^{ij} = \psi \delta^{ij} + \frac{1}{a^2} \left( \sigma_{,ij} - \frac{1}{3}
\delta_{ij} \Delta \sigma \right)+ \frac{\im}{4}\epsilon^{ijk} \omega_{,k}\, .
\eer
Here, we have introduced the factors $1/a^2$ for future convenience, $\phi$ and $\psi$
are the trace parts, $\chi$ and $\sigma$ are the scalar parts of the traceless symmetric
parts, and $\theta$ and $\omega$ are the scalar parts of the anti-symmetric parts of the
self-dual and anti-self-dual perturbations, respectively. It is easy to see that
diffeomorphism (\ref{diffeo}) can be used to set the scalars $\sigma$ and $\omega$ to
zero, while an SO(3) gauge transformation can be used to set to zero the scalar $\theta$.
Thus, we are led to consider the following ``physical'' scalar perturbations:
\beq\label{perturb-scalar}
\delta B^i = \delta B^{ij}\Sigma^j_0 + \phi \Sigma^i_0 + \psi \bar{\Sigma}^i_0\, ,
\eeq
where we have introduced a convenient notation for the non-metric part
\beq\label{perturb-s-B}
\delta B^{ij}= \frac{1}{a^2} \left( \chi_{,ij} - \frac{1}{3} \delta_{ij} \Delta \chi
\right) \, .
\eeq
In other words, the scalar perturbation (\ref{perturb-scalar}) is a sum of a non-metric
perturbation $\delta B^{ij} \Sigma^j_0$ and the usual metric perturbation
\beq\label{perturb-s-m}
\delta \Sigma^i = \phi \Sigma^i_0 + \psi \bar{\Sigma}^i_0 \, .
\eeq

The potential $\chi$ can, in general, be complex, while $\phi$ and $\psi$ are real. The
potential $\chi$ does not have a counterpart in GR, while the two potentials $\phi$ and
$\psi$ have a very simple relation to the standard GR potentials appearing in the scalar
perturbation of the metric:
\beq
ds^2 = a^2  \left[ \left(1 + 2\Phi_\GR \right) d \eta^2 - \left(1 - 2\Psi_\GR \right)
\sum_i \left( dx^i \right)^2 \right] \, .
\eeq
An elementary calculation shows the relation between $\Phi_\GR$, $\Psi_\GR$ and the
potentials $\phi$, $\psi$ in (\ref{perturb-scalar}):
\beq\label{PP-pp}
\Phi_\GR = \frac{3\psi + \phi}{2}, \qquad \Psi_\GR = \frac{\psi - \phi}{2}\, .
\eeq

\subsection{Vector sector}

The most general perturbation of the vector type can be reduced to
\beq\label{perturb-vec}
\delta B^i = 2 \left[ \zeta^{(i,j)} \Sigma^j_0 + \xi^{(i,j)} \bar{\Sigma}^j_0 \right] \,
,
\eeq
where the vector potentials $\zeta^i$ and $\xi^i$ are transverse, $\zeta^i{}_{,i} =
\xi^i{}_{,i} = 0$, and $\zeta^i$ may be complex. Alternatively, it is a sum of a
non-metric perturbation $\delta B^{ij} \Sigma^j_0$ with
\beq\label{perturb-v-B}
\delta B^{ij} = 2 \zeta^{(i,j)}
\eeq
and a metric one:
\beq\label{perturb-v-m}
\delta \Sigma^i = 2 \xi^{(i,j)} \bar{\Sigma}^j_0\, .
\eeq

\subsection{Tensor sector}

Perturbations of the tensor type that correspond to gravitons are given by
\beq\label{perturb-tens}
\delta B^i = \frac{1}{a^2}\chi^{ij}\Sigma^j + \rho^{ij} \bar{\Sigma}^j \, ,
\eeq
where both $\chi^{ij}$ and $\rho^{ij}$ are symmetric, traceless and transverse $\left(
\chi^{ij}{}_{,i} = \rho^{ij}{}_{,i}=0 \right)$ matrices, and $\chi^{ij}$ may be complex
while $\rho^{ij}$ is real (see, however, Appendix). Again, the perturbation is a sum of a non-metric one
\beq\label{perturb-t-B}
\delta B^{ij} = \frac{1}{a^2}\chi^{ij}
\eeq
and a metric one
\beq\label{perturb-t-m}
\delta \Sigma^i = \rho^{ij}\bar{\Sigma}^j_0 \, .
\eeq

\section{Scalar perturbations}
\label{sec:scalar}

\subsection{Curvature}

In this section, we consider the important case of scalar perturbations in detail. First
of all, we compute the connections and then (linearized) curvatures for the perturbations
described by $\phi$, $\psi$, and $\chi$. To compute the perturbation of the
$B$-compatible connection for a perturbation $\delta B^i$, one has to solve the algebraic
equations $\cD_0(\delta B^i) + \epsilon^{ijk} \delta A^j\wedge \Sigma^k_0=0$, where
$\Sigma^i_0$ is the background two-form. The curvature of an infinitesimal connection
$\delta A^i$ is then given by $F^i(\delta A)=\cD_0 \delta A^i$, where $\cD_0$ is the
covariant derivative with respect to the background connection. The computations are
straightforward, but rather lengthy. The expressions for the connection components are:
\ber
\delta A^i_\phi &=& \frac{\im}{2} \phi_{,i} d\eta + \frac{1}{2} \left( \im
\phi' \delta_{ij} - \epsilon_{ijk} \phi_{,k} \right) dx^j\, , \\
\delta A^i_\psi &=& \frac{3 \im}{2} \psi_{,i} d\eta - \frac{1}{2} \left[\im \frac{
\left(a^4 \phi \right)'}{a^4} \delta_{ij} - \epsilon_{ijk}\phi_{,k} \right] dx^j\, , \\
\delta A^i_\chi &=& - \frac{2 \im}{3a^2} \Delta \chi_{,i} d\eta - \frac{1}{3}
\epsilon_{ijk} \Delta \chi_{,k} dx^j - \frac{\im}{a^3}\left( a\chi_{,ij} - \frac{a}{3}
\Delta\chi \delta_{ij}\right)'dx^j \, .
\eer

The expressions for the curvatures, decomposed into their self-dual and anti-self-dual
parts, are
\ber
\cD \delta A^i_\phi &=& \frac{1}{4a^2}\left( \phi''- \Delta \phi + 2\cH
\phi'\right)\Sigma^i_0 \nonumber \\ && {} + \frac{1}{4a^2} \left[ - 2\phi_{,ij} +
\delta_{ij}(\phi''+\Delta\phi) + 2i\epsilon_{ijk}\left(\frac{\phi_{,k}}{a}\right)'a-
2\cH\phi'\delta_{ij} \right] \bar{\Sigma}^j_0 \, , \\
\cD \delta A^i_\psi &=& - \frac{1}{4a^2}\left[ 4\psi_{,ij} +
\delta_{ij}(\psi''-\Delta\psi)+ 4\delta_{ij}\left( \cH' \psi + 2 \cH^2 \psi + \frac{3}{2}
\cH \psi' \right) \right] \Sigma^j_0 \nonumber \\ && - \frac{1}{4a^2} \left[ 2 \psi_{,ij}
+ \delta_{ij} (\psi'' + \Delta \psi) + 4 \delta_{ij} \left( \cH' \psi - 2 \cH^2 \psi +
\frac{1}{2} \cH \psi' \right) \right. \nonumber \\ && \left. {}
+ 2 \im \epsilon_{ijk}
\frac{(a^3\psi_{,k})'}{a^3} \right]
\bar{\Sigma}^j_0 \, , \\
\cD \delta A^i_\chi &=& - \frac{1}{2a^4} \left[ \left(
\partial_\eta^2 - 2\cH
\partial_\eta - \Delta- 4\cH^2 \right) \left( \chi_{,ij} - \frac{1}{3} \delta_{ij}
\Delta\chi \right) \right] \Sigma^j_0 \nonumber \\
&& - \frac{1}{2a^4} \left[ \left(\partial_\eta^2 - \frac{1}{3} \Delta \right) \left(
\chi_{,ij}-\frac{1}{3}\delta_{ij}\Delta\chi \right) \right] \bar{\Sigma}^i_0 \nonumber
\\ && {} + \frac{\im}{3a^4} \epsilon_{ijk} \Delta\chi_{,k}' \bar{\Sigma}^j_0
+\frac{2}{9a^4}\Delta^2\chi \bar{\Sigma}^i_0 - \delta B^{ij}F^j_0,
\eer
where $\delta B^{ij}$ in the last formula is given by (\ref{perturb-s-B}), and $F^i_0$ is
the background curvature (\ref{F-background}). The differential operators in brackets act
only on $\chi$ but not on the background forms $\Sigma^i_0$.

\subsection{Field equations}

Now we can substitute the expressions for the curvatures obtained above into the
linearized field equations (\ref{lin-feqs*}). For the scalar sector, the sources are
\beq
\delta T = \delta \rho - 3\delta p, \qquad \delta T^{ij} = (\delta\rho + \delta p)
\delta^{ij} + \frac{2 \im}{a}\epsilon^{ijk}\delta u_{,k} - 3 \left( \delta \cS_{,ij} -
\frac{1}{3}\delta_{ij} \Delta\delta \cS\right) \, ,
\eeq
where we have introduced a convenient normalization of the scalar part of the velocity
perturbation $\delta u$. The resulting equations are most usefully separated into the
symmetric traceless, antisymmetric, and trace parts of the matrices arising in front of
the background two-forms $\Sigma^i_0$ and $\bar{\Sigma}^i_0$. Let us first analyze the
symmetric trace-free parts. After removing the directional derivatives and multiplying by
$a^2$ in the self-dual sector, and by $2a^2$ in the anti-self-dual sector, the self-dual
and anti-self-dual parts give, respectively,
\ber
&& \psi + \frac{1}{2a^2}
\left( \chi'' - 2\cH \chi' - \Delta\chi  - 4\cH^2 \chi\right) = 4\pi G \kappa \chi \, , \\
\label{eq-1} && \phi + \psi + \frac{1}{a^2} \left(\chi'' - \frac{1}{3} \Delta\chi
\right)=-12\pi G a^2 \delta\cS \, .
\eer
It is more convenient, however, to consider these equations as those determining the
usual relativistic potentials $\Psi_\GR$ and $\Phi_\GR$; see (\ref{PP-pp}). We have
\ber
\label{psigr} \Psi_{\rm GR} &:=& \frac{\psi-\phi}{2} = 4\pi G\kappa \chi +
\frac{1}{a^2}\left( \frac{1}{3}\Delta\chi +\cH \chi' + 2\cH^2\chi \right)+ 6\pi G a^2
\delta\cS\, , \\
\label{phigr} \Phi_{\rm GR} &:=& \frac{\phi+3\psi}{2} = 4\pi G\kappa \chi + \frac{1}{a^2}
\left( -\chi'' + \frac{2}{3}\Delta\chi + \cH\chi' + 2\cH^2\chi\right)- 6\pi G a^2
\delta\cS \, . \qquad
\eer
The difference between the relativistic potentials is given by
\beq \label{difference}
\Psi_{\rm GR} - \Phi_{\rm GR} = \frac{1}{a^2} \left( \chi'' - \frac13 \Delta \chi \right)
+ 12 \pi G a^2 \delta\cS \, .
\eeq
Thus, even in the absence of shear, the relativistic potentials $\Psi_\GR$ and $\Phi_\GR$
are not equal to each other in our theory, which is typical of modified gravity.

The antisymmetric part is only present in the anti-self-dual sector. After removing the
directional derivative and multiplying by $\im a^2$, one obtains
\beq\label{eq-3}
\Psi_\GR' + \cH\Phi_\GR - \frac{1}{3a^2}\Delta\chi' = 4\pi G a \delta u \, ,
\eeq
which, after solving for the perturbations $\Psi_\GR$, $\Phi_\GR$, and $\chi$, will
determine the scalar part of the 4-velocity perturbation. Finally, the trace parts of the
arising equations are most convenient in their schematic linear combinations
$\Sigma-\bar{\Sigma}$ and $3\Sigma+\bar{\Sigma}$. Using the equations of motion for the
background, after some simple algebra, we get
\ber\label{eq-4}
&& \Delta\Psi_{\rm GR} - 3\cH\Psi_{\rm GR}'-3\cH^2\Phi_{\rm GR} -
\frac{1}{3a^2} \Delta^2\chi = 4\pi G a^2 \delta\rho\, , \\
&& \Psi_{\rm GR}''+ 3\cH\Psi_{\rm GR}'+ (2\cH'+\cH^2)\Psi_{\rm GR} +
\frac{1}{3}\Delta(\Phi_{\rm GR}-\Psi_{\rm GR}) + \cH(\Phi_{\rm GR}-\Psi_{\rm GR})'
\nonumber \\ && {} +(2\cH'+\cH^2)(\Phi_{\rm GR}-\Psi_{\rm GR}) - \frac{1}{9a^2}
\Delta^2\chi = 4\pi G a^2 \delta p\, .  \label{eq-5}
\eer
Thus, the arising equations are exactly like in GR apart from the presence of simple
additional terms containing $\chi$.

\subsection{Conservation equations}

The following standard conservation equations hold in the theory under consideration:
\ber\label{conservation-0}
&& \delta\rho' + 3 \cH (\delta\rho + \delta p) = 3(\rho + p) \Psi_\GR' + \frac{1}{a}
\Delta \delta u\, , \\
\label{conservation-i} && \delta u' + 3\cH \delta u = a \delta p + a(\rho + p) \Phi_\GR +
a \Delta \delta \cS \, .
\eer
This can be verified directly using equations (\ref{eq-1}), (\ref{psigr}), and
(\ref{eq-3}) above. Note that the field $\chi$, responsible for the modification, does
not appear in these equations. This is of no surprise as the stress-energy is conserved
in this theory in the usual way.

\subsection{Equations for the ``non-metricity'' scalar $\chi$}

At this point, the usual GR procedure would be to use the fact that $\Phi_\GR = \Psi_\GR$
(in the absence of shear) and obtain a system of equations for one of the potentials. In
our case, this is not possible because any such equation will involve contributions
containing the nonmetric scalar $\chi$. However, it is possible to obtain a closed system
of equations for this scalar. Thus, we solve for $\Phi_\GR$ and $\Psi_\GR$ in terms of
the new potential $\chi$ and deal with differential equations for $\chi$. The resulting
equations are not simple, but it is worth presenting them here to show that their
structure is similar to the corresponding equations in GR. Thus, we now substitute $\psi
+ \phi = \Phi_\GR - \Psi_\GR$, $\Psi_\GR$, and $\Phi_\GR$ from (\ref{eq-1}),
(\ref{psigr}), and (\ref{phigr}), respectively, into (\ref{eq-4}) and (\ref{eq-5}), and,
after some lengthy algebra, obtain
\ber\label{constr}
4\pi G a^2 \delta \rho &=& \Delta( 4\pi G\kappa \chi) - 3\cH\left(4\pi G\kappa\chi
\right)'-3 \cH^2 \left(4\pi G\kappa\chi\right) \nonumber \\ && {} + \frac{2}{a^2}\left[
\cH^2\Delta \chi - \frac{3\cH}{2}\chi' \left(\cH' + \cH^2 \right) + 3 \cH^2 \left( \cH^2
- 2\cH' \right)
\chi\right] \nonumber \\
&& {} +6\pi G a^2\left[ \Delta \delta\cS - 3\cH \left( \delta \cS' + \cH \delta \cS
\right )\right]\, , \\
4\pi G a^2 \delta p &=& \left(4\pi G\kappa\chi \right)'' + 3 \cH \left( 4\pi G\kappa \chi
\right)' + \left(2\cH' + \cH^2 \right) \left(4\pi G\kappa \chi \right) \nonumber \\
&& {} + \frac{2}{a^2}\left[ \cH^2 \chi'' + \frac{1}{3} \Delta\chi \left(\cH'-\cH^2\right)
+ \frac{1}{2}\chi' \left(\cH'' + 7\cH'\cH - 3\cH^3 \right) \right.  \nonumber \\
&& \left. \phantom{{} + \frac{2}{a^2}\Bigl[} {} + \chi \left( 2 \cH'' \cH +
2 \cH'{}^2 - 2\cH^2 \cH' - \cH^4 \right) \right] \nonumber \\
&& {} + 6\pi G a^2\left[ \delta \cS'' + 5 \cH \left(\delta \cS'+ \cH \delta \cS \right) -
\frac{2}{3} \Delta \delta \cS\right]\, . \label{dynamical}
\eer

Similarly, substituting $\Psi_\GR$ and $\Phi_\GR$ into (\ref{eq-3}), we get
\ber
4\pi G a \delta u = \left(4\pi G\kappa\chi \right)' + \cH \left(4\pi G\kappa\chi \right)
+ \frac{1}{a^2} \left[ \chi' \left(\cH' + \cH^2 \right) + 2\chi \cH \left(2\cH' - \cH^2
\right) \right] \nonumber \\ {} + 6\pi G a^2 \left(\delta \cS'+ \cH \delta \cS\right) \,
. \label{velocity}
\eer
This equation determines the scalar part of the velocity perturbation once the
perturbations $\chi$ and $\delta \cS$ have been found.

Now, introducing the speed of sound $c_s$ via
\beq
\delta p = c_s^2 \delta \rho + \tau \delta {\mathcal E} \, ,
\eeq
we can write a single dynamical equation for $\chi$ with the entropy perturbation $\delta {\mathcal E}$
as the source. It reads:
\ber
4\pi G a^2 \tau \delta {\mathcal E} &=& \left(4\pi G\kappa\chi \right)'' - c_s^2\Delta
\left(4\pi G\kappa \chi \right) + 3\cH (1+c_s^2) \left(4\pi G\kappa \chi \right)'
\nonumber \\ && +
\left[2\cH'+\cH^2 \left( 1 + 3 c_s^2 \right)\right] \left(4\pi G\kappa\chi\right) \nonumber \\
&& {} + \frac{2}{a^2}\left(  \cH^2 \chi'' + \frac{1}{3}\Delta\chi \left[\cH'-\cH^2 \left(
1 + 3 c_s^2 \right) \right] \right. \nonumber \\ && \left. \phantom{\frac{2}{a^2}\Bigl(}
{} + \frac{\chi'}{2} \left[ \cH'' + \cH' \cH \left( 7 + 3
c_s^2 \right) + 3 \cH^3 \left( c_s^2 - 1 \right) \right] \right. \nonumber \\
&& \left. \phantom{\phantom{\frac{2}{a^2}\Bigl(}} {} + \chi \left[ 2\cH'' \cH + 2
\cH'{}^2 - 2 \cH^2 \cH' \left( 1 - 3 c_s^2 \right) - \cH^4 \left( 1 + 3 c_s^2 \right)
\right] \right)
\nonumber \\
&& {} + 6\pi G a^2 \left[ \delta \cS'' + \left( 5 + 3 c_s^2 \right) \cH \left( \delta
\cS' + \cH \delta \cS \right) - \left( \frac{2}{3} + c_s^2 \right) \Delta \delta \cS
\right] \, . \label{main-dynamical}
\eer
In the absence of shear ($\delta \cS = 0$), this dynamical equation can be solved for
$\chi$, after which one can compute all other quantities of interest, such as the
relativistic potentials and perturbations of energy density and pressure.

\subsection{New relativistic potential}

The obtained complicated equations for $\chi$ can be considerably simplified by
introducing a new quantity
\beq\label{Phi}
\Phi := \left( 4\pi G\kappa + \frac{2\cH^2}{a^2} \right) \chi \, .
\eeq
One rationale for considering precisely this combination is the identity that follows
directly from (\ref{constr}) and (\ref{velocity}):
\beq\label{dm-Phi}
4\pi G a^2 \left(\delta \rho+ \frac{3\cH}{a}\delta u - \frac{3}{2} \Delta \delta
\cS\right) = \Delta \Phi\, .
\eeq
Precisely the same identity holds in the case of GR with $\Phi$ on the right-hand side of
this equation being the relativistic potential $\Phi_{\rm GR}$, which suggests that the
quantity $\Phi$ should play the role of the main relativistic potential in our theory.
Below we shall see that this expectation is realized.

Anticipating the role that $\Phi$ is going to play, it is illuminating to rewrite the
main dynamical equation (\ref{main-dynamical}) in terms of $\Phi$. Most of the terms in
this equation combine into the usual GR-type equation for the potential $\Phi$. Few terms
remain, however, and these are the contributions due to non-metricity. It is most
convenient to write them in terms of $\chi$. We get:
\ber\label{Phi-dynamical}
4\pi G a^2 \tau \delta {\mathcal E} &=& \Phi'' - c_s^2\Delta \Phi + 3\cH \left( 1 + c_s^2
\right) \Phi'+ \left[2\cH' + \cH^2 \left( 1 + 3 c_s^2 \right) \right] \Phi \nonumber \\
&& {} + 6\pi G a^2 \left[ \delta \cS'' + 5\cH \left(\delta \cS'+ \cH \delta \cS \right) -
\frac{2}{3} \Delta \delta \cS \right] \nonumber \\ && {} - \frac{8\pi
G}{3}(\rho+p)\Delta\chi - 4\pi G \left( p'-c_s^2 \rho' \right)\chi' \, .
\eer
Here the part which does not contain $\chi$ is the standard GR dynamical equation for the
potential $\Phi$, while the part containing $\chi$ is the modification due to
non-metricity. In obtaining (\ref{Phi-dynamical}), we used the background field
equations. Let us also give an expression for the constraint (\ref{constr}) in terms of
the potential $\Phi$. Again using the background equations, we obtain
\beq\label{Phi-constr}
4\pi G a^2 \delta \rho = \Delta \Phi - 3\cH \Phi' - 3\cH^2 \Phi + 6\pi G a^2\left[ \Delta
\delta\cS - 3\cH \left( \delta \cS' +\cH \delta \cS \right) \right] - 4\pi G\rho' \chi'
\, .
\eeq
The last term here is the contribution due to non-metricity. One can also give an
equation for the velocity perturbation in terms of the new potential:
\beq\label{Phi-vel}
4\pi G a\delta u = \Phi'+\cH\Phi + 6\pi G a^2 \left( \delta \cS' +\cH \delta \cS \right)
- 4\pi G (\rho + p) \chi' \, .
\eeq
Finally, we will also need an expression for equation (\ref{dynamical}) in terms of the
new potential $\Phi\,$:
\ber\label{Phi-dyn-p}
4\pi G a^2 \delta p = \Phi'' + 3\cH \Phi'+ \left(2\cH' + \cH^2 \right) \Phi + 6\pi G a^2
\left[ \delta \cS'' + 5\cH \left( \delta \cS' + \cH \delta \cS \right) - \frac{2}{3}
\Delta \delta \cS \right] \nonumber \\ {} - \frac{8\pi G}{3} (\rho + p) \Delta \chi -
4\pi G p' \chi' \, . \qquad
\eer

\subsection{Applicability limits of the theory} \label{sec:limits}

As we have discussed above, our classical theory is applicable in the case of small
deviations from the ``metric'' behavior, which is quantitatively expressed as
(\ref{strong-coupl}). Since we have (\ref{H-Lambd}) and, in the linear theory, according
to (\ref{perturb-s-B}),
\beq
H^{ij} \approx \delta B^{ij} = \frac{1}{a^2} \left( \chi_{,ij} - \frac{1}{3} \delta_{ij}
\Delta \chi \right)\,,
\eeq
we have the following condition:
\beq \label{applic}
\frac{k^2}{a^2} | \chi | = \frac{k^2}{a^2}\,  \left| \frac{\Phi}{4 \pi G \kappa + 2 \cH^2
/ a^2} \right| \ll 1 \, .
\eeq
This condition is not stronger than the usual condition of smallness of density contrast:
\beq \label{usuapplic}
\left| \frac{\delta \rho + 3 (\cH/a) \delta u - \frac32 \Delta \cS}{\rho} \right| \ll 1
\quad \Longleftrightarrow \quad \frac{2 k^2}{3 \cH^2 - \Lambda_0 a^2}\, |\Phi| \ll 1 \, ,
\eeq
provided the denominator $4 \pi G \kappa + 2 H^2$ in (\ref{applic}) does not approach
zero sufficiently closely.  We have to assume that this never happens and that always
\beq \label{positive}
2 \pi G \kappa + \frac{\cH^2}{a^2} > 0 \, .
\eeq
Note that this condition is also necessary in order that the change of variable
(\ref{Phi}) be non-degenerate.  Recalling the definition (\ref{kappa}) of $\kappa$, we
obtain a constraint on the parameter $\beta$ or $g$ by considering a dust-dominated
universe: $ \beta > - 4/3$, or $g > -1$. Considering the inflationary universe with $p
\approx - \rho$, we get a stronger condition $\beta > - 1/3$, or $g > 0$.  Finally,
considering the ``stiff'' matter with $p = \rho$ (which is realized, e.g., by the
kinetically-dominated regime of a scalar field), we have the constraint $\beta < 2/3$, or
$g < 1$. Thus, the physical values of $g$ or, respectively, $\beta = g - 1/3$ lie in the
domain
\beq \label{betacon}
0 < g < 1 \, , \qquad - \frac13 < \beta < \frac23 \, .
\eeq

In a similar way, we obtain a constraint on the parameter $\gamma$ by considering a
universe which is dominated by the cosmological constant: $\gamma > - 4/3$. An even
stronger condition can be obtained by considering a radiation-dominated universe. We then
have:
\beq \label{gammacon}
\gamma > 0\, .
\eeq
We, therefore, assume that inequalities (\ref{betacon}) and (\ref{gammacon}) are
satisfied by a large margin.  In particular, the quantity $\beta$ cannot be much larger
than unity by absolute value.

\section{Evolution of scalar perturbations} \label{sec:evolution}

\subsection{Inflation} \label{sec:inflation}

The primordial spectrum of perturbations is obtained in the inflationary paradigm.
Therefore, we start with application of our theory to the simplest model of inflation
based on a single inflaton field $\varphi$.  We assume that the inflaton has the usual
coupling to the metric defined by the $B$-field. Then its action can be written as
\beq
S[\varphi] = \int d^4x \sqrt{- {\rm det}\left( g_{\mu\nu} \right)}\left[ g^{\mu\nu}
\partial_\mu \varphi
\partial_\nu \varphi - V(\varphi)\right] \, ,
\eeq
where $g_{\mu\nu}$ is the metric defined by $B$.

The background field equations for $\varphi$ take the usual form
\beq\label{infl-background}
\cH^2 = \frac{8\pi G}{3} \left[ \frac12 \varphi'^2 + a^2 V(\varphi) \right] +
\frac{\Lambda_0 a^2}{3} \, , \qquad \cH^2 - \cH' = 4\pi G \varphi'^2 \, ,
\eeq
where, as before, the prime denotes the derivative with respect to the conformal time.

The equations for perturbations can be obtained from the general equations (\ref{constr})
and (\ref{velocity}) by substituting appropriate expressions for $\delta \rho$ and
$\delta u$:
\beq\label{infl-rho-u}
\delta \rho \to \left( \frac{\varphi'}{a} \right)^2 \left[ \left(
\frac{\delta\varphi}{\varphi'} \right)' - 2\cH \frac{\delta\varphi}{\varphi'} - \Phi_\GR
\right]\, , \qquad \delta u \to \frac{\varphi'}{a} \delta\varphi\, .
\eeq
From these expressions, we find
\beq\label{infl-1}
\delta \rho + \frac{3\cH}{a}\delta u = \left( \frac{\varphi'}{a} \right)^2 \left[ \left(
\frac{\delta\varphi}{\varphi'} \right)' + \cH \frac{\delta\varphi}{\varphi'} - \Phi_\GR
\right]\, .
\eeq
To obtain a dynamical equation for perturbations, it remains to express this quantity in
terms of the potential $\Phi$ using (\ref{dm-Phi}). For this purpose, we first use
equation (\ref{eq-3}) to express $\Phi_\GR$ in terms of the potential $\Psi_\GR$:
\beq\label{infl-2}
\cH \Phi_\GR = \left( \cH^2 - \cH' \right) \frac{\delta\varphi}{\varphi'} +
\frac{1}{3a^2} \Delta \chi' - \Psi_\GR' \, ,
\eeq
where we have used expression (\ref{infl-rho-u}) for $\delta u$ as well as the background
field equations (\ref{infl-background}). Substituting (\ref{infl-2}) into (\ref{infl-1})
and once again using the background field equations, we can write equation (\ref{dm-Phi})
as follows:
\beq\label{infl-3}
\frac{\cH^2-\cH'}{\cH} \left( \cH \frac{\delta\varphi}{\varphi'} + \Psi_\GR \right)' -
\frac{\cH^2-\cH'}{3a^2\cH} \Delta\chi' = \Delta\Phi\, ,
\eeq
where, as before, $\Phi$ is given by (\ref{Phi}).

Now we use expressions (\ref{velocity}) and (\ref{psigr}), respectively, for $\delta u$
and $\Psi_\GR$ in terms of $\chi$ to transform the expression in the brackets of
(\ref{infl-3}) into the following form:
\beq\label{infl-4}
\cH \frac{\delta\varphi}{\varphi'} + \Psi_\GR = \frac{\cH^2/a^2}{\cH^2-\cH'}\left(
\frac{a^2 \Phi}{\cH}\right)' + \frac{1}{3a^2} \Delta \chi\, ,
\eeq
where we have used the definition (\ref{Phi}) of the potential $\Phi$. Combining
equations (\ref{infl-3}) and (\ref{infl-4}), we see that the term containing
$\Delta\chi'$ cancels. The remaining term with $\Delta\chi$ can be converted to a term
with $\Delta \Phi$ by using the relation (\ref{Phi}) between $\chi$ and $\Phi$. The final
equation for the potential $\Phi$ is
\beq \label{conform}
\frac{\cH^2-\cH'}{\cH} \left[ \frac{\cH^2/a^2}{\cH^2-\cH'}\left( \frac{a^2
\Phi}{\cH}\right)' \right]' = \left[ 1 + \frac{\cH^2 - \cH'}{3 \left( \cH^2 + 6 \pi G a^2
\kappa \right)} \right] \Delta \Phi\, .
\eeq
Let us also present it in terms of the derivatives with respect to the physical time:
\beq \label{inflat}
\frac{\dot{H}}{H} \partial_t \left[ \frac{H^2}{a\dot{H}}\partial_t \left( \frac{a\Phi}{H}
\right)\right] = \left[ 1 - \frac{\dot{H}}{3 \left( H^2 + 6 \pi G \kappa \right) }\right]
\frac{\Delta \Phi}{a^2}\, .
\eeq
Apart from the term on the right-hand side proportional to $\dot{H}$, this is the usual
equation for the evolution of the relativistic potential during inflation.  Note that the
denominator $H^2 + 6 \pi G \kappa $ in the last term on the right-hand side of
(\ref{inflat}) is positive by virtue of constraints (\ref{betacon}).

The quantity $6 \pi G \kappa$ in the case under consideration is given by the expression
\beq \label{kapin}
6 \pi G \kappa = 2 \pi G (3 g - 1) (\rho - 3 p) + \frac{3}{4 \ell^2} - \Lambda_0 = ( 3g -
1) \left( 3 H^2 + \frac32 \dot H \right) + \frac{3}{4 \ell^2} - 3 g \Lambda_0 \, .
\eeq
As we have already noted in Sec.~\ref{sec:lambda}, the equations for perturbations, in
general, are not invariant under the simultaneous change
\beq \label{shift1}
\Lambda_0 \to \Lambda_0 - \lambda \, , \qquad V (\varphi) + \frac{\lambda}{8 \pi G} \, ,
\eeq
which would be the case in general relativity.  However, they are invariant after an
additional change
\beq \label{shiftel}
\frac{1}{\ell^2} \to \frac{1}{\ell^2} - 4 g \lambda \, .
\eeq
In particular, such a transformation will arise in the case where the parameter $\ell^2$
is fixed to be $\ell^2 = 1/4 g \Lambda_0$.

During the inflationary epoch, we have $|\dot{H}| \ll H^2$, and one can neglect the term
containing $\dot{H}$ on the right-hand side of (\ref{inflat}). Thus, we can argue that
the evolution of inflation-generated perturbations is unchanged in the theory under
consideration, and we can take the standard flat spectrum for the potential $\Phi$ as
initial conditions. This conclusion for the generated spectrum can be justified as
follows. Equations (\ref{dm-Phi}) and (\ref{infl-1}) imply the relation
\beq \label{match}
\Delta \Phi + 4 \pi G \varphi'^2 \Phi_\GR = \frac{4 \pi G \varphi'^2}{a} \left( \frac{a
\delta \varphi}{\varphi'} \right)' \, .
\eeq
On spatial scales smaller than the Hubble scales, with the comoving wavenumber $k$
satisfying $k^2 \gg \cH^2$, we can neglect the self-gravity of the scalar field and
quantize the inflaton perturbation $\delta \varphi$ on the background of a homogeneous
and isotropic inflationary universe, obtaining the standard spectrum for the modes
$\delta \varphi_{\bf k}$.  Then we can use equation (\ref{match}) to match the quantities
$\Phi$ and $\delta \varphi$ on small (sub-Hubble) spatial scales during inflation. In
doing this, the term $4 \pi G \varphi'^2 \Phi_\GR$ in (\ref{match}) can be neglected
(just as it is the case with a similar term in GR) since, according to (\ref{phigr}), it
is estimated as
\beq
4 \pi G \varphi'^2 \Phi_\GR = 4 \pi G \varphi'^2  \left[\Phi + \cO \left( \cH^{-1}
\right) \Phi' + \cO \left( \cH^{-2} \right) \Phi'' + \cO \left( \cH^{-2} \right) \Delta
\Phi \right] \, ,
\eeq
and is small in view of the condition $k^2 \gg \cH^2 \gg 4 \pi G \varphi'^2$. Thus, the
quantity $\Phi$ on small scales will acquire the standard amplitude with scale-invariant
spectrum and then will evolve according to equation (\ref{inflat}), which, as we noted,
differs negligibly from the standard inflationary equation for the relativistic
potential.

In the regime of very long wavelengths, we can neglect all terms with Laplacians.  In
this case, equation (\ref{match}) implies the relation
\beq
\Phi_\GR \approx \frac{1}{a} \left( \frac{a \delta \varphi}{\varphi'} \right)' = \left(
\frac{\delta \varphi}{\varphi'} \right)' + \frac{\cH \delta \varphi}{\varphi'}\, .
\eeq
Substituting this into (\ref{infl-rho-u}), we get
\beq \label{deltarho}
\delta \rho \approx - 3 \cH  \left( \frac{\varphi'}{a} \right)^2 \frac{\delta
\varphi}{\varphi'} = - \frac{3 \cH}{a^2} \varphi' \delta \varphi \, .
\eeq
Equations (\ref{psigr}) and (\ref{phigr}) with Laplacians neglected give
\beq \label{long}
\Psi_\GR \approx \Phi + \frac{\cH}{a^2} \chi' \, , \qquad \Phi_\GR \approx \Phi -
\frac{1}{a^2} \chi'' + \frac{\cH}{a^2} \chi' \, .
\eeq
Then, according to (\ref{infl-background}) and (\ref{infl-2}), we have
\beq \label{deltaphi}
4 \pi G \varphi' \delta \varphi \approx \cH \Phi_\GR + \Psi'_\GR \approx \cH \Phi + \Phi'
- 4 \pi G \varphi'^2 \chi' \approx \cH \Phi \, ,
\eeq
where, in the last approximation, we have neglected the two terms $\Phi' - 4 \pi G
\varphi'^2 \chi'$ compared to $\cH \Phi$, which is legitimate during inflation. Comparing
(\ref{deltarho}) and (\ref{deltaphi}), we get the usual relation
\beq
\frac{\delta \rho}{\rho} \approx - 2 \Phi \, ,
\eeq
valid on super-Hubble spatial scales during inflation.  Note that, in the same
inflationary approximation, we have $\Phi_\GR \approx \Psi_\GR \approx \Phi$ on
super-Hubble spatial scales.

\subsection{Evolution with a generic equation of state}

In this subsection, we would like to demonstrate that, on asymptotically large spatial
scales, the relativistic potential $\Phi$ defined in (\ref{Phi}) behaves just as in
general relativity, including the cases where the universe experiences slow or rapid
transitions between epochs with different effective equations of state.  To see this,
consider equation (\ref{Phi-dynamical}) in the case of vanishing shear $\cS$ for a system
described by a generic equation of state $p(\rho)$, which can interpolate between
different regimes. Note that this assumption implies vanishing of the entropy
perturbation, $\delta {\mathcal E} = 0$, as well as of the last term in equation
(\ref{Phi-dynamical}), which equation then takes the form
\beq \label{simple}
\Phi'' - c_s^2\Delta \Phi + 3\cH \left( 1 + c_s^2 \right) \Phi'+ \left[2\cH' + \cH^2
\left( 1 + 3 c_s^2 \right) \right] \Phi - \frac{8\pi G}{3}(\rho+p)\Delta\chi = 0 \, .
\eeq
By the standard change of function (see \cite[Sec.~7.3]{Mukhanov})
\beq
\Phi = u \exp \left[ - \frac32 \int \left( 1 + c_s^2 \right) \cH d \eta \right]  \propto
( \rho + p )^{1/2} u \, ,
\eeq
equation (\ref{simple}) is transformed to
\beq \label{equ}
u'' - c_\eff^2 \Delta u - \frac{\vartheta''}{\vartheta} u = 0 \, ,
\eeq
where
\beq
\vartheta \equiv \frac{1}{a} \left[ \frac23 \left( 1 - \frac{\cH'}{\cH^2} \right)
\right]^{-1/2} = \frac{1}{a} \left( \frac{\rho + \rho_\Lambda}{\rho + p} \right)^{1/2} \,
,
\eeq
\beq \label{ceff-u}
c_\eff^2 = c_s^2 + \frac{2 (\rho + p)}{3 \beta (\rho - 3 p) + 3 \gamma \rho_\Lambda + 4
\rho} \, ,
\eeq
and we have used definition (\ref{Phi}) and neglected the cosmological constant
contribution in the denominator.

The long-wave solution of (\ref{equ}) is obtained when the term with Laplacian is
neglected:
\beq
u (\eta) \approx C \vartheta (\eta) \int_{\eta_0}^\eta \frac{d \eta'}{\vartheta^2
(\eta')} \, ,
\eeq
where $C$ and $\eta_0$ are constants of integration.  In this regime, the quantity
\beq
\zeta = \vartheta^2 \left( \frac{u}{\vartheta} \right)' \propto \Phi + \frac23 \frac{\rho
+ \rho_\Lambda}{\rho + p} \left(\Phi +  \frac{1}{\cH} \Phi' \right)
\eeq
is conserved.

Equation (\ref{equ}) can be recast in the form
\beq \label{equ1}
\left[ \vartheta^2 \left( \frac{u}{\vartheta} \right)' \right]' = c_\eff^2 \vartheta^2
\Delta \left( \frac{u}{\vartheta} \right) \, ,
\eeq
which, in particular, shows that the quantities
\beq
\frac{u}{\vartheta} = a \left( \frac{\rho + p}{\rho + \rho_\Lambda} \right)^{1/2} u
\propto \frac{a^2}{\cH} \Phi
\eeq
and
\beq
\vartheta^2 \left( \frac{u}{\vartheta} \right)' - \frac{\vartheta^2}{3 \cH} \Delta \left(
\frac{u}{\vartheta} \right) \propto \Phi + \frac23 \frac{\rho + \rho_\Lambda}{\rho + p}
\left(\Phi +  \frac{1}{\cH} \Phi'  - \frac{1}{3 \cH^2} \Delta \Phi \right)
\eeq
 should remain continuous during a rapid transition between different equations of
state. Since $a$ and $\cH$ are obviously continuous in this case, it follows that $\Phi$
must also be continuous.

The relations derived in this subsection imply that the infla\-tionary and
post-infla\-tionary evolution of the potential $\Phi$ on large spatial scales will
reproduce those of general relativity. The only modification as compared with GR occurs
on small spatial scales, where oscillations in (\ref{equ}) proceed with effective speed
of sound (\ref{ceff-u}), different from $c_s^2$.  During some periods in the cosmological
evolution, this difference can be small and lead to negligible effect in (\ref{equ}). In
this case, characterized by
\beq
\frac{2 (\rho + p)}{3 \beta (\rho - 3 p) + 3 \gamma \rho_\Lambda + 4 \rho} \ll c_s^2 \, ,
\eeq
we will say that perturbations evolve in the regime of {\em general relativity\/} (GR) on
all scales. In the opposite case, when the difference between $c_\eff^2$ and $c_s^2$ is
essential, we will say that they evolve in the regime of {\em modified gravity\/}.

From (\ref{psigr}) and (\ref{phigr}), one can obtain relations between the relativistic
potentials on small scales:
\ber \label{psimod}
\Psi_{\rm GR} &=& \displaystyle \Phi + \frac{1}{a^2} \left( \frac13 \Delta \chi + \cH
\chi' \right) \approx \left[ 1 - \frac{k^2}{12 \pi G a^2 \kappa + 6 \cH^2} \right] \Phi
\, , \quad k^2 \gg \cH^2 \, , \\ \nonumber \\
\Phi_{\rm GR} &=& \displaystyle \Phi + \frac{1}{a^2} \left( \frac23 \Delta \chi - \chi''
+ \cH \chi' \right) \approx \left[ 1 + \frac{\left( 3 c_\eff^2 - 2 \right) k^2}{12 \pi G
a^2 \kappa + 6 \cH^2} \right] \Phi \, , \quad k^2 \gg \cH^2 \, . \qquad \label{phimod}
\eer
In the regime of general relativity, where $2 \pi G \kappa \gg \cH^2/a^2$, we have
\beq \label{potgr}
\Psi_{\rm GR} \approx \left[ 1 - \frac{2k^2 \ell^2}{3 a^2} \right] \Phi \, , \quad
\Phi_{\rm GR} \approx \left[1 + 2 \left(c_\eff^2 - \frac23 \right)\frac{k^2 \ell^2}{a^2}
\right] \Phi \, , \quad k^2 \gg \cH^2 \, .
\eeq
In deriving the above relations, we have taken into account that, in the high-frequency
regime under consideration, $\chi'' \approx c_\eff^2 \Delta \chi$, so that the term
proportional to $\chi''$ in (\ref{phigr}) gives a contribution of the same order as
$\Delta\chi$. We see that, even in the GR regime, the usual GR relation $\Phi_\GR
=\Psi_\GR$ is violated for physical wave numbers $k/a > 1/\ell$, i.e., on scales smaller
than the scale of deformation $\ell$.

\subsection{Radiation domination}\label{sec:rad-dom}

The evolution of scalar perturbations in a universe filled with fluid are most easily
analyzed using the system of equations (\ref{Phi-dynamical}), (\ref{Phi-constr}). For
simplicity, we set the shear to zero and restrict ourselves to adiabatic perturbations,
$\delta {\mathcal E}=0$.

At the stage of radiation domination, we have $a=a_0\eta$, so that $\cH=1/\eta$,
$\cH'=-\cH^2$, and $\cH''= 2 \cH^3$. The speed of sound at this stage is $c_s^2=1/3$.
Equation (\ref{Phi-dynamical}) then simplifies to
\beq\label{rad-dom}
\Phi'' + \frac{4}{\eta}\Phi' - c_\eff^2 \Delta \Phi = 0 \, ,
\eeq
where
\beq \label{eff-cs}
c_\eff^2 = \frac{\gamma \rho_\Lambda +4 \rho}{3 \gamma \rho_\Lambda + 4\rho} \, ,
\eeq
and, in using equation (\ref{Phi}),  we have taken into account that $\rho = 3p$, hence,
$\kappa = \gamma \rho_\Lambda = \mbox{const}$. Equation (\ref{rad-dom}) has the form of
the usual equation for a radiation-dominated universe apart from the fact that the
effective speed of sound became time-dependent.

In a very early universe, $\rho \gg \gamma \rho_\Lambda$, and the effective speed of
sound (\ref{eff-cs}) becomes equal to the speed of light. This is the modified-gravity
regime. In the course of time, if the quantity $\gamma \rho_\Lambda$ becomes dominant
over $\rho$, the effective speed of sound (\ref{eff-cs}) turns to its standard value
$1/\sqrt{3}$. This is the GR regime.  As usual, the limit of GR is obtained by sending
$\gamma \to \infty$.

From the constraint (\ref{Phi-constr}), we find the expression for energy-density
perturbations during this stage:
\beq\label{drho-raddom}
4\pi G a^2 \delta \rho = \Delta\Phi - 3\cH \Phi' \left( \frac{3 \gamma \rho_\Lambda}{3
\gamma \rho_\Lambda + 4 \rho} \right) - 3\cH^2 \Phi \left[ 1- \frac{4(4\rho)^2}{(3 \gamma
\rho_\Lambda + 4 \rho)^2} \right] \, ,
\eeq
where we have used the fact that $\kappa = \gamma \rho_\Lambda = \mbox{const}$ during
radiation domination and that $\rho' = - 4 \cH \rho$. Again, as $\gamma \to \infty$, this
equation takes the standard GR form.

As usual, the qualitative behavior of solutions of (\ref{rad-dom}) is easy to understand
in the two asymptotic regimes: $k \eta \ll 1$, corresponding to modes with wavelength
larger than the Hubble radius, and $k\eta \gg 1$, corresponding to modes that already
entered the Hubble radius. In the first case, the term $\Delta \Phi$ can be neglected,
and $\Phi = \mbox{const}$ is a non-decreasing solution. In this case, the last term in
equation (\ref{drho-raddom}) is dominant and gives $a^4 \delta \rho \propto \Phi$ with
coefficient of proportionality being a function of time. Note that the usual
general-relativity relation $\delta\rho/\rho = -2\Phi$ is no longer valid in our theory.
Indeed, this will hold only in the GR regime, where $\rho \ll \gamma \rho_\Lambda$, while
in the early universe we may have $\rho \gg \gamma \rho_\Lambda$. Thus, for $k \eta \ll
1$, we have
\beq \label{rad-relation}
\frac{\delta \rho}{\rho} \approx \left\{
\begin{array}{rl}
6 \Phi \, , &\rho \gg \gamma \rho_\Lambda  \ \mbox{(modified-gravity regime)} \, ,\\
- 2 \Phi \, , &\rho \ll \gamma \rho_\Lambda \ \mbox{(GR regime)} \, .
\end{array} \right.
\eeq
As a consequence, the long-wave gauge-invariant energy-density perturbation
$\delta\rho/\rho$ passes through zero for $\gamma \rho_\Lambda \sim \rho$, which is an
interesting phenomenon connected with the gauge choice in this model.

Note that the standard relation
\beq
\frac{\delta \rho}{\rho} \approx - 2 \Phi_\GR
\eeq
remains to be true in both regimes, in view of relation (\ref{long}) between the
potentials $\Phi$ and $\Phi_\GR$ valid in the long-wave approximation.  Since it is the
quantity $\Phi$ which remains constant on large scales during radiation domination, both
$\delta \rho / \rho$ and $\Phi_\GR$ change in the transition between the regimes of
modified gravity and GR.  Relations (\ref{long}) in this case read
\ber
\Psi_\GR &\approx& \left[ 1 + \frac{2 (4 \rho)^2}{( 3 \gamma \rho_\Lambda + 4 \rho)^2}
\right] \Phi \approx \left\{
\begin{array}{rl}
3 \Phi \, , &\rho \gg \gamma \rho_\Lambda  \ \mbox{(modified-gravity regime)} \, ,
\\
\Phi \, , &\rho \ll \gamma \rho_\Lambda \ \mbox{(GR regime)} \, .
\end{array} \right. \\  \nonumber \\
\Phi_\GR &\approx& \left[ 1 + \frac{4 \left( 9 \gamma \rho_\Lambda - 4 \rho \right) (4
\rho)^2}{( 3 \gamma \rho_\Lambda + 4 \rho)^3} \right] \Phi \approx \left\{
\begin{array}{rl}
- 3 \Phi \, , &\rho \gg \gamma \rho_\Lambda  \ \mbox{(modified-gravity regime)} \, ,
\\
\Phi \, , &\rho \ll \gamma \rho_\Lambda \ \mbox{(GR regime)} \, .
\end{array} \right. \ \qquad
\eer

For modes with $k\eta\gg 1$, the term $\Delta \Phi$ in (\ref{rad-dom}) is important. The
dominant solution of (\ref{rad-dom}) oscillates with frequency determined by the
effective speed of sound (\ref{eff-cs}) and with a decaying amplitude $\propto \eta^{-2}
c_\eff^{-1/2} \propto a^{-2} c_\eff^{-1/2}$. In this case, the first term on the
right-hand side of (\ref{drho-raddom}) is dominant, and the relation between $\delta \rho
/ \rho$ and the potential $\Phi$ is the same as in GR. The energy-density perturbation
$\delta\rho/\rho$ is now an oscillating function with frequency determined by
(\ref{eff-cs}) and with amplitude $\propto c_\eff^{-1/2}$, which is constant both in the
modified and GR epochs and growing during the transition.  The relativistic potentials on
such small scales are described by the general equations (\ref{psimod})--(\ref{potgr}).

Thus, as in GR, perturbations do not grow during radiation domination. This is true both
for the modified-gravity epoch in early universe and for the GR epoch at late times.
However, the relation between the energy density fluctuation $\delta\rho/\rho$ and the
main relativistic potential $\Phi$ on the super-Hubble scales is different at these
epochs, as shown by (\ref{rad-relation}).  The effective speed of sound (\ref{eff-cs}),
which determines oscillations on sub-Hubble scales, is also different at different
epochs. During the transition occurring for $\gamma \rho_\Lambda \sim \rho$, if it is
realized at the radiation-dominated stage, the long-wave gauge-invariant energy-density
perturbation $\delta\rho/\rho$ passes through zero.

Perturbations at the radiation-dominated stage are studied in greater detail in
Sec.~\ref{sec:power} below.

\subsection{Matter domination} \label{sec:matter}

At this stage, we have $p = \delta p =0$, $a (\eta) \propto \eta^2$, and equation
(\ref{Phi-dynamical}) simplifies to
\beq\label{mat-dom}
\Phi'' + \frac{6}{\eta}\Phi' - c_\eff^2 \Delta \Phi = 0 \, ,
\eeq
where
\beq \label{effm-cs}
c_\eff^2 = \frac{2 \rho}{3\kappa + 4\rho} \, .
\eeq
In this case, $\kappa = \beta \rho + \gamma \rho_\Lambda$ is not constant [$\beta$ and
$\gamma$ are defined in (\ref{bg})]. On sufficiently large spatial scales, $k \ll \cH /
c_\eff$, we have $\Phi \approx \rm{const}$ while, on small scales, $k \gg \cH / c_\eff$,
the potential $\Phi$ oscillates with decreasing amplitude $\propto \eta^{-3}
c_\eff^{-1/2} \propto a^{-3/2} c_\eff^{-1/2}$ and with effective speed of sound given by
(\ref{effm-cs}). This behavior is quite different from the case of general relativity,
where the corresponding relativistic potential remains constant at all spatial scales.

To determine the behavior of the density contrast, consider the quantity
\beq\label{delta-m}
\delta_m := \frac{\delta \rho+ 3(\cH/a)\delta u}{\rho} \, ,
\eeq
which is related to $\Phi$ through (\ref{dm-Phi}):
\beq\label{delta-chi}
4\pi G a^2 \rho\, \delta_m = \Delta \Phi \, .
\eeq
We see that, on large spatial scales, $k \ll \cH / c_\eff$, the density contrast grows
like in GR: $\delta_m \propto a$, while, on small spatial scales, $k \gg \cH / c_\eff$,
it {\em oscillates\/} with a slightly increasing amplitude
$\propto a^{-1/2} c_\eff^{-1/2}\propto a^{1/4}$ and frequency
determined by the effective speed of sound in (\ref{effm-cs}).

It is useful to obtain a closed equation describing the evolution of $\delta_m$.  Using
the conservation equations (\ref{conservation-0}) and (\ref{conservation-i}) as well as
our assumption $p = \delta p =0$, we get
\beq
\rho \delta_m' = 3 \rho \left( \Psi_\GR' + \cH \Phi_\GR \right) + \frac{1}{a}\Delta
\delta u + \frac{3}{a} \left( \cH' - \cH^2 \right) \delta u \, .
\eeq
Taking into account equation (\ref{eq-3}) and the background equation $\cH^2 - \cH' = 4
\pi G a^2 (\rho + p)$, we see that the terms proportional to $\delta u$ cancel, so that
\beq\label{m-1}
\delta_m' = \frac{1}{a^2} \Delta \chi' + \frac{1}{a\rho} \Delta \delta u \, .
\eeq

Taking the second derivative of $\delta_m$, we obtain
\beq\label{m-2}
\delta_m'' = \frac{1}{a^2} \left(\Delta\chi'' - 2 \cH \Delta \chi' \right) + \Delta
\Phi_\GR - \frac{\cH}{a\rho} \Delta \delta u \, ,
\eeq
where we have used the conservation equation (\ref{conservation-i}) once again. The
following identity is a direct consequence of (\ref{eq-1}) and (\ref{psigr}):
\beq\label{m-3}
\Delta \Phi_\GR + \frac{1}{a^2} \Delta \left(\chi''-\cH \chi' \right) =
 \frac{2}{3a^2}\Delta^2\chi + \left( 4\pi G\kappa + \frac{2\cH^2}{a^2}\right)\Delta \chi \, .
\eeq
Combining (\ref{m-1})--(\ref{m-3}) and using the relation (\ref{delta-chi}) between
$\delta_m$ and $\Delta\chi$, we have
\beq
\delta_m''+\cH\delta_m'-  \frac{2}{3a^2} \Delta^2\chi= 4\pi G a^2\rho \delta_m \, .
\eeq
Finally, using the relation (\ref{Phi}) between $\Phi$ and $\chi$ as well as equation
(\ref{delta-chi}), we obtain
\beq
\delta_m''+\cH\delta_m' -  \frac{2\rho}{3\kappa+4\rho} \Delta\delta_m = 4\pi G a^2\rho
\delta_m \, .
\eeq
It is illuminating to rewrite the above equation in terms of the physical time:
\beq\label{dm}
\ddot{\delta}_m + 2H \dot{\delta}_m -\frac{2\rho}{3\kappa+4\rho}
\frac{\Delta\delta_m}{a^2} = 4\pi G \rho \delta_m \, ,
\eeq
where $H=\dot{a}/a$ is the physical Hubble parameter. Equation (\ref{dm}) allows for a
straightforward passage to the GR limit: one just has to send $\kappa\to\infty$. For
finite $\kappa$, we get an extra term that leads to oscillations on sub-Hubble scales, an
effect absent in general relativity.

In view of the definition $\kappa=\beta \rho +\gamma \rho_\Lambda$, we can write:
\beq \label{ceff-m}
c_\eff^2 = \frac{2/3}{\beta + 4/3 + \gamma \rho_\Lambda/\rho} \, .
\eeq
Thus, at the early epoch with $\rho\gg \gamma \rho_\Lambda$ (if this condition is
realized during matter domination), we have oscillations with effective speed of sound
$c_\eff^2=2/(3\beta+4)$. If $\gamma \gg 1$, then, as the energy density decreases with
the expansion, one may enter into the regime $\rho_\Lambda \ll \rho \ll \gamma
\rho_\Lambda$ in which the usual GR behavior is recovered in a broad range of spatial
scales. Note that the physical spatial scale on which oscillations take place
$k/a> \cH/c_{\eff}a \sim a^{-3/2} (\gamma\rho_\Lambda/\rho)^{1/2}\sim 1/\ell$ remains
constant during the matter domination era.

\subsection{Lambda domination}

At later epochs in its evolution, the universe is dominated by the cosmological constant.
It is thus necessary to consider the evolution of matter perturbations in such a
universe. The scale factor in a Lambda-dominated universe is given by
\beq
a (\eta) = - \frac{1}{\eta} \sqrt{\frac{3}{\Lambda_0}} \, ,
\eeq
so that $\cH = - 1 / \eta$ and $\cH' = 1 / \eta^2 = \cH^2$.  Equation
(\ref{Phi-dynamical}) in this approximation reads
\beq \label{lambda-dom}
\Phi'' - \frac{3}{\eta} \Phi' + \frac{3}{\eta^2} \Phi - c_\eff^2 \Delta \Phi = 0 \, ,
\eeq
with the effective speed of sound
\beq
c_\eff^2 = \frac{4 \pi G \rho }{6 \pi G \kappa + 3 \cH^2 / a^2 } = \frac{
2\rho}{(3\beta+4)\rho + (3\gamma+4)\rho_\Lambda } \sim
\frac{2}{3\gamma+4} (\rho/\rho_\Lambda) \ll 1\, ,
\eeq
where we have taken into account that $\cH^2 / a^2 = \Lambda_0 / 3 + 8 \pi G \rho / 3$,
$\kappa = \beta \rho + \gamma \rho_\Lambda$, and the $\Lambda$-domination
condition $\rho_\Lambda\gg \rho$. Equation (\ref{lambda-dom}) in this
case differs from the general-relativistic analog only by the term with $c_\eff^2 \Delta
\Phi$.

On sufficiently large scales, $k \ll \cH / c_\eff $, the evolution proceeds as in the
case of general relativity, with the dominant mode
\beq
\Phi \propto \eta \propto a^{-1} \, .
\eeq
The evolution of the density contrast is given by equation (\ref{Phi-constr}):
\beq
4 \pi G a^2 \delta \rho = \Delta \Phi - 4 \pi G \rho' \chi' = \Delta \Phi - \frac92
c_\eff^2 \cH^2 \Phi \, .
\eeq
One can see that the evolution is the same as in general relativity, $\delta \rho / \rho
\propto a \Phi = \mbox{const}$ in the range of scales $c_\eff \cH \ll k \ll \cH /
c_\eff$.  However, the lower boundary in this case ($c_\eff^2 \ll 1$) corresponds to the
scales well beyond the Hubble radius, which are unobserved in our universe.  Thus, we may
conclude that the long-wave perturbations, with $k \ll \cH / c_\eff $, evolve in the GR
regime.

On small scales, $k \gg \cH / c_\eff $, equation (\ref{lambda-dom}) describes
oscillations with adiabatically decreasing amplitude $\propto a^{-3/2} c_\eff^{-1/2}
\propto a^{-3/4}$ (in the universe with pressureless matter). In this case, for $c_\eff^2
\ll 1$, we have
\beq
4 \pi G a^2 \delta \rho \approx \Delta \Phi \, ,
\eeq
so that the density contrast $\delta \rho / \rho$ oscillates with amplitude $\propto
a^{-1/2} c_\eff^{-1/2} \propto a^{1/4}$, similar to what happens in the regime of
matter domination.

The threshold comoving frequency which distinguishes between these two types of behavior
is growing as $k_\Lambda := \cH / c_\eff \propto a^{5/2}$, and so the physical threshold
scale $a / k_\Lambda$ decreases from its value $\sim \ell$ at the matter-dominated stage
as $a/k_\Lambda \propto a^{-3/2}$.

\section{Effects of the modification of gravity on the power spectrum} \label{sec:power}

In this section, we give a qualitative analysis of the effect of modification of gravity
on the linear evolution of matter and radiation density perturbations in the adiabatic
mode.

First of all, as we have seen in Sec.~\ref{sec:inflation}, the spectrum on super-Hubble
scales generated during inflation is practically the same as in the usual inflationary
theory based on general relativity.  The subsequent evolution of the spectrum depends on
the epoch at which the condition $\rho = \gamma \rho_\Lambda$ is reached, i.e., it
depends on the value of the parameter $\gamma$.

Dealing with the post-inflationary evolution, we consider a two-component universe filled
by non-interacting radiation and dark matter (we ignore the small contribution from the
baryons at the radiation-dominated epoch). Using the basic system of equations
(\ref{conservation-0}), (\ref{conservation-i}) and (\ref{Phi})--(\ref{Phi-dyn-p}), it is
possible to derive a closed system of second-order differential equations for the two
convenient variables, namely, the potential $\Phi$ and the entropy perturbation
\beq \label{s-m}
s_m := \frac34 \delta_r - \delta_m \, ,
\eeq
where, in this subsection, we have made the notation $\delta_r = \delta \rho_r / \rho_r$
and $\delta_m = \delta \rho_m / \rho_m$ for radiation and dark matter, respectively.  One
of these two equations is just equation (\ref{Phi-dynamical}), and the combined system in
Fourier space reads [equation (\ref{sys-s}) is derived in the appendix]
\ber \label{sys-phi}
&& \Phi'' + 3 \cH \left(1 + c_s^2 \right) \Phi' + \left[ k^2 c_\eff^2 + 2 \cH' + \cH^2
\left( 1 + 3 c_s^2 \right) \right] \Phi = 4 \pi G a^2 c_s^2 \rho_m s_m \, , \\
&& \frac{1}{3 c_s^2} s_m'' + \cH s_m' + \frac{k^2}{4} \frac{\rho_m}{\rho_r} s_m =
\frac{k^4}{16 \pi G a^2 \rho_r} \Phi \, . \label{sys-s}
\eer
System (\ref{sys-phi}), (\ref{sys-s}) differs from the corresponding system in general
relativity only by the presence of $c_\eff^2$ instead of $c_s^2$ in one place in
(\ref{sys-phi}), where $c_s^2$ and $c_\eff^2$ are respectively given by
\beq \label{sound}
c_s^2 = \frac{1}{3} \left( 1 + \frac{3 \rho_m}{4 \rho_r} \right)^{-1} \, , \qquad
c_\eff^2 = c_s^2 + \frac{2 (\rho + p)}{3 \kappa + 4 \rho} \, .
\eeq

Once a solution to (\ref{sys-phi}) and (\ref{sys-s}) is found, the density contrasts can
be determined using (\ref{s-m}) and (\ref{Phi-constr}):
\ber \label{deltar}
\delta_r &=& - \frac{3 c_s^2}{4 \pi G a^2 \rho_r} \left[ 3 \cH \Phi' + \left( k^2 + 3
\cH^2 \right) \Phi \right] + \frac{4 \cH}{a^2} \chi' + 3 c_s^2 \frac{\rho_m}{\rho_r} s_m
\, , \\
\delta_m &=& - \frac{9 c_s^2}{16 \pi G a^2 \rho_r} \left[ 3 \cH \Phi' + \left( k^2 + 3
\cH^2 \right) \Phi \right] + \frac{3 \cH}{a^2} \chi' - 3 c_s^2 s_m \, , \label{deltam}
\eer
where $\chi$ is expressed through $\Phi$ by (\ref{Phi}).  These equations formally differ
from their general-relativistic counterparts only by the terms containing $\chi$.

At the radiation-dominated epoch, equations (\ref{sys-phi}) and (\ref{sys-s}) take the
simple form
\ber \label{rad-phi}
&& \eta^2 \Phi'' + 4 \eta \Phi' + k^2 c_\eff^2 \eta^2 \Phi = \frac12 \Upsilon \eta s_m \, , \\
&& \eta^2 s_m'' + \eta s_m' + \frac{k^2}{4} \Upsilon \eta^3 s_m = \frac{k^4}{6} \eta^4
\Phi \, , \label{rad-s}
\eer
respectively.  Here $\Upsilon$ is a dimensionless parameter defined by
\beq
\Upsilon = \left( \frac{8 \pi G}{3} \rho_\eq a_\eq^2 \right)^{1/2} \, , \qquad \Upsilon
\eta \ll 1 \, ,
\eeq
and $a_\eq$ and $\rho_\eq$ are the scale factor and radiation density at the
radiation-matter equality: $\rho_m = \rho_r = \rho_\eq$.

If the parameter $\gamma$ is such that the equality $\rho = \gamma \rho_\Lambda$ is
reached during inflation, then the evolution of the spectrum at the radiation-dominated
stage is the same as in general relativity.  This follows from equations (\ref{sys-phi})
and (\ref{sys-s}), which, in this case, coincide with those of general relativity at the
radiation-dominated stage. Since inflation typically ends at energy densities $\rho_{\rm
infl} \sim 10^{-12}/ 4 \pi G^2$, and since
\beq
\gamma \rho_\Lambda \approx \frac{1}{8 \pi G \ell^2} \, , \qquad \gamma \gg 1 \, ,
\eeq
we have the estimate for $\ell$ in this case:
\beq
\ell < 10^{6} \ell_P \sim 10^{-27}\, \mbox{cm} \, ,
\eeq
where $\ell_P = G^{1/2} \simeq 10^{-33}\,\mbox{cm}$ is the Planck length.  Such small
values of $\ell$ make the theory practically indistinguishable from the general
relativity (which, we remember, is obtained in the limit $\ell \to 0$).

Assume that $\gamma \gg 1$ is such that the condition $\rho = \gamma \rho_\Lambda$ takes
place at the radiation-dominated epoch. Let us determine the range of values of the
parameter $\ell$ for which this is the case. The energy density $\rho$ at the
matter-radiation dominated stage is expressed as
\beq
\rho \simeq \frac{3 H_0^2}{8 \pi G} \left[ \Omega_m (1 + z)^3 + \Omega_r (1 + z)^4
\right] = \frac{3 H_0^2}{8 \pi G} \Omega_m (1 + z)^3 \left[ 1 + \frac{1 + z}{1 + z_{\rm
eq}} \right] \, ,
\eeq
where $z_{\rm eq}$ is the redshift of matter-radiation equality, $H_0$ is the current
value of the Hubble parameter, and $\Omega_r$ and $\Omega_m$ are the radiation and matter
density parameters, respectively. When making numerical estimates, for definiteness, we
will use the recently determined values of these parameters \cite{Komatsu:2010fb}:
\beq \label{params}
\Omega_m h_{75}^2 \approx 0.24 \, , \qquad z_{\rm eq} \simeq 3 \times 10^3 \, ,
\eeq
where $h_{75} = H_0 / 75 \, \mbox{km}/\mbox{s Mpc} \approx 1$.

The condition $\gamma \rho_\Lambda = \rho$ defines the redshift $z_\ell\,$:
\beq \label{zel}
1 + z_\ell = \left( \frac{1 + z_{\rm eq}}{3 H_0^2 \Omega_m \ell^2} \right)^{1/4} \, ,
\qquad z_\ell \gg z_\eq \, ,
\eeq
and the condition $z_\ell > z_{\rm eq}$ implies
\beq \label{ell-rad}
\ell < \left[ 6 \Omega_m H_0^2 (1 + z_{\rm eq})^3 \right]^{-1/2} \simeq 20\, \mbox{kpc}
\, .
\eeq

For the modes that enter the Hubble radius well in the GR regime, i.e., at $z < z_\ell$,
the spectrum of perturbations will not be modified by the end of the radiation-dominated
stage as compared to general relativistic cosmology, which can be seen as follows. Before
the Hubble-radius crossing, the terms with $k$ in system (\ref{sys-phi}) and
(\ref{sys-s}) can be dropped, after which this system coincides with the corresponding
system in GR. After the transition to GR regime, when $c_\eff^2$ becomes equal to
$c_s^2$, the system again coincides with that of general relativity for all scales $k$,
as was already discussed in Sec.~\ref{sec:rad-dom}. Therefore, the behavior of the
solutions $\Phi$ and $s_m$ is general-relativistic during the whole evolution.  The
non-standard terms with $\chi$ in equations (\ref{deltar}) and (\ref{deltam}) also become
unimportant in the GR regime.  Hence, the density contrasts are expressed through the
basic functions $\Phi$ and $s_m$ also in a general-relativistic way. Therefore, their
amplitudes and phases will not be modified compared to the general-relativistic
expressions.

The boundary value of the comoving spatial scale $\lambda_\ell = a_0 / k_\ell$ that we
were talking about is given by the condition that the corresponding wave crosses the
Hubble radius at redshift $z_\ell$. We have
\beq \label{kell}
k_\ell = \cH_\ell = \frac{a_\ell}{\sqrt{3}\, \ell} \, ,
\eeq
from which, using (\ref{zel}), we get
\beq \label{lambdal}
\lambda_\ell = \frac{a_0}{k_\ell} = \sqrt{3} (1 + z_\ell) \ell \simeq \lambda_\eq \left(
\frac{1 + z_\ell}{1 + z_\eq} \right) \left( \frac{\ell}{20\, \mbox{kpc}} \right) \simeq
100 \left( \frac{\ell}{20\,\mbox{kpc}} \right)^{1/2}\, \mbox{Mpc} \, ,
\eeq
where $\lambda_\eq = a_0 / k_\eq \simeq 100\, \mbox{Mpc}$ is the comoving spatial scale
corresponding to Hubble-radius crossing at matter-radiation equality, and, we remember,
this estimate works for $\ell < 20\,\mbox{kpc}$. For $\lambda
> \lambda_\ell$, or $k < k_\ell = a_0 / \lambda_\ell$, the spectrum of
perturbations will not be modified by the end of the radiation-dominated stage.

Consider now the modes that enter the Hubble radius at the modified-gravity epoch, where
$c_\eff^2 \approx 1$. Such modes satisfy $k \gg k_\ell$; their comoving spatial scales,
therefore, are considerably smaller than (\ref{lambdal}). In studying the evolution of
{\em adiabatic\/} perturbations on such scales, for not very large values of $\eta$, we
can neglect the terms with $\Upsilon$ in equations (\ref{rad-phi}) and (\ref{rad-s}).
Then, before the transition to the GR regime, we have $c_\eff^2 \approx \mbox{const} =
1$, and the exact solution of this system that describes the non-decaying mode with the
property $s_m \to 0$ as $\eta \to 0$ is given by
\ber
\Phi &=& \frac{3 \Phi_0}{(k c_\eff \eta)^2} \left( \frac{\sin k c_\eff \eta}{k c_\eff
\eta} - \cos k c_\eff \eta
\right) \, , \label{phi-ad} \\
s_m  &=&  \frac{\Phi_0}{c_\eff^4} \left[\frac{\cos k c_\eff \eta - 1}{2}  - \int_0^{k
c_\eff \eta} \frac{\cos x -1}{x} d x \right] \nonumber \\ &=& \frac{\Phi_0}{c_\eff^4}
\left[\frac{\cos k c_\eff \eta - 1}{2} + {\cal C} + \ln k c_\eff \eta - {\rm Ci} ( k
c_\eff \eta) \right]\, , \label{s-ad}
\eer
where $\Phi \approx \Phi_0 = \mbox{const}$ at $k \eta \ll 1$, ${\cal C} \approx 0.577$ is
the Euler constant, and ${\rm Ci} (x)$ is the cosine integral. We have retained the
constant $c_\eff$ in equations (\ref{phi-ad}) and (\ref{s-ad}) which, therefore, will be
applicable also to the case of general relativity, where $c_\eff^2 = c_s^2 = 1/3$.

Consider now a wave with a fixed comoving wave number $k$. It enters the Hubble
radius at the moment $\eta\sim 1/c_{\eff} k$. Thus, when we are in the modified
gravity regime the moment of Hubble-radius crossing occurs for a given wave earlier
than would be the case in GR. The oscillations in $\Phi$ thus start earlier, and the amplitude
of the gravitational potential $\Phi$ also drops more than in GR. A similar suppression
effect is also present in the matter power spectrum. From the above expression for
the entropy perturbation we learn that $s_m\propto (k \eta)^4$ for $k c_{\eff} \eta \ll 1$.
Thus, the entropy perturbation grows till the mode enters the Hubble radius where oscillations
start. Since in the modified gravity epoch the Hubble-radius entry occurs earlier, the
entropy perturbation grows considerably less then would be the case in GR and
a suppression of matter power spectrum ensues.  These effects can be estimated as follows.

Well after the Hubble-radius crossing (at $k c_{eff} \eta > 1$), we can proceed to a new function
$u = \eta^2 \Phi$, for which equation (\ref{rad-phi}) with zero right-hand side gives
\beq
u'' + \left( k^2 c_\eff^2 - \frac{2}{\eta^2} \right) u = 0 \, .
\eeq
Therefore, an approximate solution to $\Phi$ in the regime $k^2 c_\eff^2 \gg \cH^2 =
\eta^{-2}$ can be given in the WKB form
\beq \label{solu}
\Phi \propto \frac{1}{\eta^2 \sqrt{c_\eff}} \cos \int k c_\eff d \eta \, .
\eeq
Comparing this with the leading terms in (\ref{phi-ad}), (\ref{s-ad}), we conclude that
the WKB solution well after the Hubble-radius crossing can be approximated by
\ber
\Phi &\approx& - \frac{3 \Phi_0}{(k c_\eff^0 \eta)^2 } \sqrt{\frac{c_\eff^0}{c_\eff}}
\cos \int_0^\eta k c_\eff (\eta') d \eta' \, , \label{phi-wkb} \\ \nonumber \\ s_m
&\approx& \frac{\Phi_0}{2 \left( c_\eff^0 \right)^2 c_\eff^2}
\sqrt{\frac{c_\eff^0}{c_\eff}} \cos \int_0^\eta k c_\eff (\eta') d \eta' +
\frac{\Phi_0}{\left( c_\eff^0 \right)^4} \left( {\cal C} - \frac12 + \ln k c_\eff^0 \eta
\right) \, , \label{s-wkb}
\eer
where $c_\eff^0$ is the asymptotic initial value of the effective sound velocity, which
we have inserted here for comparison with the case of general relativity.  The second
term in (\ref{s-wkb}) is just a source-free solution of equation (\ref{rad-s});
therefore, we have retained its form without any modification.

We note that the same expressions (\ref{phi-wkb}), (\ref{s-wkb}) will be obtained in
general relativity with the substitution of $c_s$ for $c_\eff$ everywhere.  Therefore,
after the transition from the modified-gravity regime, where $c_\eff^2 (\eta) \approx 1$,
to the GR regime, where $c_\eff^2 (\eta) = c_s^2 (\eta)  \approx 1/3$, solution
(\ref{phi-wkb}) for the potential $\Phi$ will evolve just like in the case of general
relativity except that its phase will be shifted and the amplitude of its oscillations
will be lower by a factor of
\beq \label{factor}
c_s^{-3/2} = 3^{3/4} \approx 2.3 \, .
\eeq
This comprises the suppression by $c_s^2$ due to an earlier entry into the regime of
acoustic oscillations with decaying amplitude, as well as amplification of the amplitude
by the factor $c_s^{-1/2}$ that occurs in the transition from the modified-gravity to the
GR regime.

The amplitudes of the radiation and matter energy densities on small scales
($k c_{\eff} \gg \cH$)
will be given, according to (\ref{deltar}), (\ref{deltam}), by the expressions
\ber
\delta_r \approx - \frac{3 c_s^2 k^2}{4 \pi G a^2 \rho_r} \Phi &=& \frac{6 \Phi_0
c_s^2}{\left( c_\eff^0 \right)^2} \sqrt{\frac{c_\eff^0}{c_\eff}} \cos \int_0^\eta k
c_\eff (\eta') d \eta' \, , \label{r-wkb} \\ \nonumber \\
\delta_m \approx \frac34 \delta_r - 3 c_s^2 s_m &\approx& \frac{9 \Phi_0 c_s^2}{2 \left(
c_\eff^0 \right)^2} \sqrt{\frac{c_\eff^0}{c_\eff}} \left( 1 - \frac{1}{3 c_\eff^2}
\right) \cos \int_0^\eta k c_\eff (\eta') d \eta' \nonumber \\ && {} - \frac{3 c_s^2
\Phi_0}{\left( c_\eff^0 \right)^4} \left( {\cal C} - \frac12 + \ln k c_\eff^0 \eta
\right) \, . \label{m-wkb}
\eer

We see that the amplitude of $\delta_r$, compared with its general-relativistic
counterpart, after transition to GR regime (where $c_\eff^2 = c_s^2 = 1/3$), will be
lower by the same factor (\ref{factor}), which will affect the CMB pattern on small
angular scales.  If one wishes to exclude this modification in the CMB spectrum at
multipoles $l \lesssim 2500$, corresponding to the values reached by ACBAR
\cite{Reichardt:2008ay}, then one needs to impose the following constraint on the
parameter $\ell$:
\beq \label{ell-cmb}
\frac{l_{\rm rec}}{l_\ell} = \frac{\lambda_\ell}{\lambda_{\rm rec}} =
\frac{\lambda_\ell}{\lambda_\eq} \frac{\lambda_\eq}{\lambda_{\rm rec}} \simeq 0.4 \left(
\frac{\ell}{20\, \mbox{kpc}} \right)^{1/2} \lesssim \frac{200}{2500} \quad \Rightarrow
\quad \ell \lesssim 1\, \mbox{kpc} \, .
\eeq
Here, $l_{\rm rec} \approx 200$ is the characteristic multipole number corresponding to
the Hubble radius crossing at recombination, and $\lambda_{\rm rec} = a_0 / k_{\rm rec}
\approx 250\, \mbox{Mpc}$ is the corresponding comoving length.

The amplitude of matter perturbation on small scales under consideration is given by
(\ref{m-wkb}).  In general relativity, we would always have $c_\eff^2 = c_s^2 = 1/3$, and
the oscillatory term would always be absent.  In the present case of modified gravity,
the matter density contrast, apart from the usual monotonic evolution described by the
last term in (\ref{m-wkb}), also exhibits oscillations.  However, these oscillations
cease after transition to the GR regime (where $c_\eff^2 = c_s^2 = 1/3$), and the matter
density contrast is then described by the usual monotonic term. Its amplitude is smaller
than that of general-relativistic expression by the factor $c_s^4 = 1/9$, although the
argument in the logarithm is larger by a factor of $\sqrt{3}$.  We thus conclude that, in
the eventual power spectrum $P(k) \propto | \delta_m (k) |^2$ of dark matter, there will
be an additional suppression factor on small scales generated during the
radiation-dominated stage:
\beq \label{transfer}
\frac{P_{\rm mod} (k)}{P_{\Lambda{\rm CDM}} (k)} = \left[\frac{{\cal C} - \frac12 + \ln k
\eta_\eq }{9 \left({\cal C} - \frac12 + \ln \frac{k \eta_\eq}{\sqrt{3}}\right)} \right]^2
\approx \frac{1}{80} \, , \qquad k \gg k_\ell > k_\eq \, ,
\eeq
where $P_{\rm mod} (k)$ and $P_{\Lambda{\rm CDM}} (k)$ are the power spectra in the
modified gravity theory under consideration and in general relativity, respectively,
$\eta_\eq$ is the conformal time at the matter-radiation equality, and $k_\ell$ and its
corresponding comoving length are given, respectively,  by (\ref{kell}) and
(\ref{lambdal}).  Such a strong suppression will place additional constraint on the value
of $\ell$.  In order that it does not disturb the observed power above the comoving
spatial scales of $\sim 1\, \mbox{Mpc}$, one would require $\lambda_\ell \lesssim 1\,
\mbox{Mpc}$, or
\beq \label{finres}
\ell \lesssim 2\, \mbox{pc}\, .
\eeq
On the other hand, suppression of the power of linear perturbations on comoving scales
below $1\, \mbox{Mpc}$ may be interesting from the viewpoint of the missing-satellite
problem (see \cite{Kravtsov:2009gi}).

During matter domination, there will be an additional modification of power caused by
oscillations on scales $k > k_m = \cH / c_\eff \simeq a / \sqrt{2} \ell$ with slightly
increasing amplitude $\delta_m \propto a^{1/4}$ (see Sec.~\ref{sec:matter}). The
corresponding comoving length scale is
\beq \label{lambdam}
\lambda_m = \frac{a_0}{k_m} \simeq \sqrt{2} \ell\, (1 + z) \simeq 8\, \left( \frac{1 +
z}{1 + z_\eq} \right) \left( \frac{\ell}{2\, \mbox{pc}} \right) \, \mbox{kpc} \, .
\eeq
At present, the evolution of perturbations on such small spatial scales is non-linear,
and the linear-theory analysis of this paper is no longer applicable.  On the other hand,
it is not easy to probe the spectrum on such scales in the linear regime in the early
universe. It is thus hard to see whether the new effect of matter density oscillations
could be detected with the currently available data.

The analysis made in this section, in particular, suggests that, if the condition $\rho =
\gamma \rho_\Lambda$ took place at the matter-dominated stage, which requires values of
$\ell$ larger than those given in (\ref{ell-rad}), this would modify the power spectra of
dark matter and radiation very significantly compared to the case of general relativity,
which is important from the viewpoint of current observations. For instance, the pattern
in the CMB power spectrum would be affected because of the essential difference in the
effective speed of sound $c_\eff^2$ from the usual one $c_s^2$ in this case, and the
matter power spectrum would be suppressed on all scales because of specific oscillations
on the matter-dominated stage. Whether it is possible to fit all available observations
in this case by varying cosmological parameters requires additional investigation using
numerical methods.

Finally, we note that we have studied here only the adiabatic mode, for which $s_m \to 0$
asymptotically in the past.  For the bound (\ref{finres}) on the parameter $\ell$ that
arose in this section, the entropy mode is practically excluded, just as it is the case
in the $\Lambda$CDM model. It also remains to be seen whether one can loosen the
restrictions on the parameter $\ell$ in the presence of an entropy mode.

\section{Discussion}

The modified theory of gravity that was studied in this paper has several specific
features which make it an interesting object of investigation.  Expressed in terms of the
self-dual two-forms of the Pleba\'{n}ski formalism, the theory is local and has the same
number of degrees of freedom as general relativity.  Due to this property, the theory, in
particular, respects the analog of the Birkhoff theorem, possessing a unique
one-parameter family of (modified\,!) spherically symmetric vacuum solutions
\cite{Krasnov:2007ky}, which also turn out to be static.  Unlike some other modifications
of general relativity, such as massive gravity, it has a continuous limit to general
relativity and, therefore, may be regarded as a smooth deformation of the latter.  At the linearized
level, the modification is described by a parameter $\ell$ with dimension of length and a
dimensionless parameter $g$; they enter the fundamental potentials $V$ and $R$ defined in
(\ref{V}) and (\ref{R}), respectively. The value of $g$ is restricted to lie between zero
and unity from the requirement of absence of singularity in the equations for linear perturbations,
while the parameter $\ell$ is essentially free. The limit of general relativity is
obtained as $\ell \to 0$, irrespective of the value of $g$.

The cosmological properties of the theory also reflect the absence of new degrees of
freedom.  Thus, the dynamics of an {\em ideally\/} homogeneous and isotropic universe
described by the modified gravity under consideration is absolutely the same as in
general relativity. For a realistic universe, which is homogeneous only statistically,
there will be the problem of averaging which might lead to modification of the effective
Friedmann equations, the issue that lies beyond the scope of the present paper.

The theory of linear perturbations is modified in a very interesting way. Concerning
perturbations of the scalar type, first of all, there exists an analog of the
relativistic potential $\Phi$ in this theory, which is related to the matter
perturbations in the usual way through Eq.~(\ref{dm-Phi}). Just as in general relativity,
it is also possible to obtain a system of second-order partial differential equations
describing the evolution of this potential; see Eqs.~(\ref{Phi})--(\ref{Phi-dyn-p}). This
system is different from its counterpart in general relativity.  Its analysis, performed
in Secs.~\ref{sec:evolution} and \ref{sec:power}, reveals the following basic qualitative
features:
\begin{enumerate}
\item The inflationary theory of generation of primordial perturbation remains intact
    because corrections due to modification of gravity at the inflationary stage are
    suppressed by the small ratio $| \dot H | / H^2 \ll 1$.

\item The evolution of perturbations at a radiation-dominated stage, described by
    (\ref{rad-dom}), proceeds with a modified effective time-dependent speed of sound
    $c_\eff$, given by (\ref{eff-cs}).  The relation between the energy density and
    relativistic potential, Eq.~(\ref{drho-raddom}), is also modified as compared to
    general relativity.  For sufficiently low values of the length parameter, $\ell <
    20\, \mbox{kpc}$, given by (\ref{ell-rad}), the transition from the
    modified-gravity regime (where $c_\eff \approx 1$) to the general-relativistic
    regime (where $c_\eff \approx c_s \approx 1/\sqrt{3}$) takes place at the
    radiation-dominated stage.  The modes that enter the Hubble radius after this
    transition evolve just as in general relativity, so that the spectrum in this
    long-wave region is not modified.  The modes that enter the Hubble radius before
    this transition experience some suppression in the course of the transition; their
    amplitude drops by approximately $2.3$, see Eq.~(\ref{factor}). This will
    affect the CMB angular power spectrum on sufficiently small scales; in order that
    this spectrum remain unmodified below the multipole number $l \simeq 2500$
    reached by ACBAR \cite{Reichardt:2008ay}, the fundamental length parameter of the
    theory is further restricted to be $\ell \lesssim 1\, \mbox{kpc}$; see
    Eq.~(\ref{ell-cmb}).

    Modification of gravity during the radiation-dominated stage also has interesting
    effect on the evolution of dark-matter perturbations on small scales, which is
    described in Sec.~\ref{sec:power}.  Specifically, the matter density contrast,
    apart from the usual monotonic evolution also exhibits oscillations, described by
    the first term in (\ref{m-wkb}), that cease after transition to the GR regime
    (where $c_\eff^2 = c_s^2 = 1/3$).  Moreover, the amplitude of the monotonically
    growing part of matter perturbation is smaller than the similar expression in
    general relativity by the factor $c_s^4 = 1/9$.  This leads to a suppression of
    the eventual power spectrum of dark matter by a factor $\sim 1/80$ [see
    (\ref{transfer})] on the scales that enter the Hubble radius well before the
    transition to general-relativistic regime.  In order that this suppression does
    not disturb the observed power above the comoving spatial scales of $\sim 1\,
    \mbox{Mpc}$, one would require $\ell \lesssim 2\, \mbox{pc}$, which turns out to
    be the strongest constraint on this parameter in the present scenario. On the
    other hand, suppression of the power of linear perturbations on comoving scales
    below $1\, \mbox{Mpc}$ may be interesting from the viewpoint of the
    missing-satellite problem (see \cite{Kravtsov:2009gi}).

\item The evolution at the matter-dominated stage is qualitatively different from
    that of general relativity because equation (\ref{mat-dom}) for the relativistic
    potential still contains a nonzero effective speed of sound, given by
    (\ref{effm-cs}) or (\ref{ceff-m}). For this reason, the closed equation
    (\ref{dm}) for density contrast exhibits {\em oscillatory\/} character on
    sufficiently small scales; the corresponding comoving spatial scale is given by
    (\ref{lambdam}).  In view of the previous constraint $\ell \lesssim 2\,
    \mbox{pc}$, this behavior will only affect the comoving spatial scales smaller
    than about $8\, \mbox{kpc}$. Note, however, that, at present, the evolution of
    perturbations on such small spatial scales is non-linear, and the linear-theory
    analysis of this paper is no longer applicable.  On the other hand, it is not
    easy to probe the spectrum on such scales in the linear regime in the early
    universe.  It is thus hard to see whether the new effect of matter density
    oscillations could be detected with the currently available data.

\end{enumerate}

We should note that the above description is of qualitative character and, therefore,
needs to be substantiated by numerical integration of the derived system of exact
equations for perturbations, which will be the subject of subsequent work. We also leave
to the future work an analysis of the effect of modification on the gravitational waves,
with a preliminary description given in Appendix \ref{sec:tens}.

\acknowledgments

K.\,K.\@ was supported by an EPSRC Advanced Fellowship. Yu.\,S.\@ was supported by the
``Cosmomicrophysics'' programme and Program of Fundamental Research of the Physics and
Astronomy Division of the National Academy of Sciences of Ukraine, and by the State
Foundation for Fundamental Research of Ukraine under grant F28.2/083.

\appendix
\section*{Appendix}
\section{Derivation of equations describing the radiation--dark-matter system}

The conservation equations (\ref{conservation-0}) and (\ref{conservation-i}), written in
Fourier space for a system of matter and radiation with vanishing shear, read
\ber
& \delta_m' = \theta_m + 3 \Psi_\GR' \, , \qquad   & \delta_r' = \frac43 \theta_r + 4
\Psi_\GR' \, , \label{deltas} \\
& \theta_m' = - \cH \theta_m - k^2 \Phi_\GR \, , \qquad  & \theta_r' = - k^2 \Phi_\GR -
\frac14 k^2 \delta_r \, . \label{thetas}
\eer
Here, the variable $\theta_a$ for each component is defined as
\beq
\theta_a = - \frac{k^2 \delta u_a}{\rho_a + p_a} \, .
\eeq

Equations (\ref{Phi-constr}) and (\ref{Phi-vel}) in the case of vanishing shear can be
expressed, respectively, as
\ber
4 \pi G a^2 \left( \rho_m \delta_m + \rho_r \delta_r \right) &=& - 3 \cH \left( \Phi' +
\cH \Phi \right) - k^2 \Phi - 4 \pi G \rho' \chi' \, , \label{deltaq} \\
4 \pi G a^2 \left( \rho_m \theta_m + \frac43 \rho_r \theta_r \right) &=& - k^2
\left[\Phi' + \cH \Phi - 4 \pi G (\rho + p) \chi' \right] \, . \label{thetaq}
\eer

Differentiating the entropy perturbation $s_m = \frac34 \delta_r - \delta_m$, using
(\ref{deltas}) and (\ref{thetas}),  we get
\beq
s_m' = \theta_r - \theta_m \, , \qquad s_m'' = \cH \theta_m - \frac{k^2}{4} \delta_r \, .
\eeq

It remains to calculate the left-hand side of (\ref{sys-s}).  Using the definition
(\ref{sound}) of $c_s^2$, we have
\ber
&& \frac{1}{3 c_s^2} s_m'' + \cH s_m' + \frac{k^2}{4} \frac{\rho_m}{\rho_r} s_m =
\nonumber \\ && =  \left( 1 + \frac{3 \rho_m}{4 \rho_r} \right) \left( \cH \theta_m -
\frac{k^2}{4} \delta_r \right) + \cH \left(\theta_r - \theta_m \right) + \frac{k^2}{4}
\frac{\rho_m}{\rho_r} \left(\frac34 \delta_r - \delta_m \right) \nonumber \\
&& = \frac{3 \cH}{4 \rho_r} \left( \rho_m \theta_m + \frac43 \rho_r \theta_r \right) -
\frac{k^2}{4 \rho_r} \left( \rho_m \delta_m + \rho_r \delta_r \right) \nonumber \\
&& = {} - \frac{3 \cH}{4 \rho_r}  \frac{k^2}{4 \pi G a^2} \left[\Phi' + \cH \Phi - 4 \pi
G (\rho + p) \chi' \right] \nonumber \\ && \phantom{=\ } {} + \frac{k^2}{4 \rho_r}
\frac{1}{4 \pi G a^2} \left[ 3 \cH \left( \Phi' + \cH \Phi \right) + k^2 \Phi + 4 \pi G
\rho' \chi' \right] = \frac{k^4}{16 \pi G a^2 \rho_r} \Phi \, . \label{app-fin}
\eer
Here, in the third equality, we have used equations (\ref{deltaq}) and (\ref{thetaq}),
and, in the last equality, the conservation equation $\rho' = - 3 \cH (\rho + p)$.
Equation (\ref{app-fin}) is the desired result (\ref{sys-s}).

\section{Tensor sector}
\label{sec:tens}

\subsection{Curvatures}

To describe the linearized connections and the curvature in the tensor sector it is
very convenient to introduce the following two differential operators on the space of symmetric
traceless transverse matrices:
\beq
DX^{ij}:= X^{ik}{}_{,l}\epsilon^{jkl} + \im (X^{ij})' \, , \qquad
\bar{D}X^{ij}:= X^{ik}{}_{,l} \epsilon^{jkl} - \im (X^{ij})' \, ,
\eeq
where, as usual, the prime denotes the conformal time derivative. It is easy to see that
the operators $D$ and $\bar{D}$ transform symmetric transverse traceless matrices into
matrices with similar properties and that
\beq\label{D}
D \bar{D} = \bar{D} D = -\Box \, ,
\eeq
where we have denoted $\Box:=  -\partial_\eta^2 +\Delta$. In terms of the derivative
operators introduced, we get
\beq
\delta A_\chi^i = -\frac{1}{a^3}D(a \chi^{ij}) dx^j\, , \qquad \delta A_\rho^i = -
\frac{1}{a} \bar{D}(a \rho^{ij}) dx^j\, .
\eeq
For a linearized connection of the form $\delta A^i=A^{ij} dx^j$ with a symmetric
traceless matrix $A^{ij}$, the linearized curvature is given by the following expression:
\beq
\cD_0 \delta A^i = \frac{1}{2a} \bar{D} \left(\frac{A^{ij}}{a}\right)  \Sigma^j_0 -
\frac{1}{2a^3} D(a A^{ij}) \bar{\Sigma}^j_0\, .
\eeq
Thus, we have
\ber
\cD_0 \delta A_\chi^i &=& - \frac{1}{2a} \bar{D} \left[ \frac{D \left( a \chi^{ij}
\right)}{a^4} \right] \Sigma^j_0 + \frac{1}{2a^3}\bar{D} \left[ \frac{D \left( a
\chi^{ij} \right)}{a^2} \right] \bar{\Sigma}^j_0\, , \\
\cD_0 \delta A_\rho^i &=& - \frac{1}{2a} \bar{D} \left[ \frac{\bar  D \left( a\rho^{ij}
\right)}{a^2} \right] \Sigma^j_0 + \frac{1}{2a^3} D \bar{D} (a\rho^{ij})\,
\bar{\Sigma}^j_0\, .
\eer
Note that the derivative operators $D$ and $\bar D$ do not act on the basis two-forms in
these expressions.

\subsection{Field equations}

For tensor perturbations, there is no source on the right-hand side of (\ref{lin-feqs*}),
and we get the following simple system of equations:
\ber
&& a \bar{D} \left[ \frac{D \left( a \chi^{ij} \right)}{a^4} \right] -
\frac{1}{a^2}\left(\cH^2 + \cH' \right) \chi^{ij} + a \bar{D} \left[ \frac{\bar D \left(
a\rho^{ij} \right)}{a^2} \right] - \left(\cH^2 - \cH' \right) \rho^{ij} = 8\pi G \kappa
\chi^{ij}\, , \ \qquad  \\
&& \frac{1}{a} D \left[ \frac{D \left( a \chi^{ij} \right)}{a^2} \right] -
\frac{1}{a^2}\left(\cH^2 - \cH' \right) \chi^{ij} + \frac{1}{a} D \bar{D} (a \rho^{ij}) -
\left(\cH^2 + \cH' \right) \rho^{ij} = 0 \, .
\eer
Now expanding the derivative operators one obtains the following coupled
system of second-order differential equations:
\ber \label{tens-eq1}
\frac{1}{a^2} \left( \partial_\eta^2 -\Delta - 2\cH \partial_\eta - 4\cH^2 - 8\pi G a^2
\kappa +4\im \cH \epsilon \right) \chi^{ij} = \left( \partial_\eta^2 +\Delta +2\im
\partial_\eta \epsilon \right) \rho^{ij}\, ,
\\ a^2 \left( \partial_\eta^2 -\Delta + 2\cH \partial_\eta \right) \rho^{ij}= \left(
\partial_\eta^2 +\Delta -2\im \partial_\eta\epsilon \right) \chi^{ij}\, , \label{tens-eq2}
\eer
where $\epsilon$ is a first-order differential operator that acts on the space of
symmetric trace-free matrices via:
\beq
\epsilon X^{ij} = X^{ik}{}_{,l} \epsilon^{jkl} \, .
\eeq
We note that $\epsilon^2 = -\Delta$.
As a check, in the limit $\chi^{ij}\to 0$ such that $\kappa\chi^{ij}$ is kept finite, we
get the GR result $(\partial_\eta^2-\Delta + 2 \cH \partial_\eta) \rho^{ij} = 0$.

\subsection{Action principle and Hamiltonian analysis}

We have written the system of equations (\ref{tens-eq1}), (\ref{tens-eq2}) in the above
form to make it obvious that they can be obtained as Euler--Lagrange equations for the
following Lagrangian:
\ber\label{act-tens}
{\cal L}_{tens} &=& - \frac{a^2}{2} \rho \left( \partial_\eta^2-\Delta+2\cH\partial_\eta
\right) \rho + \chi \left( \partial_\eta^2 +\Delta +2\im \partial_\eta \epsilon \right)
\rho \nonumber \\
&& -\frac{1}{2a^2} \chi \left( \partial_\eta^2-\Delta-2\cH\partial_\eta -4\cH^2 -8\pi
G\kappa a^2 + 4\im \cH \epsilon \right) \chi\, ,
\eer
where we have dropped the internal indices for brevity. Indeed, one easily obtains both
equations using the fact that the operators $\partial_\eta^2-\Delta+2\cH\partial_\eta$,
$\partial^2_\eta$, and $\partial_\eta^2-\Delta-2\cH\partial_\eta$ are self-adjoint with
respect to the inner products given by
\beq
\langle \phi, \psi \rangle_+ = \int d\eta d^3x \, a^2  \phi \psi\, , \quad \langle \phi,
\psi \rangle_0 = \int d\eta d^3x \, \phi \psi\, , \quad \langle \phi, \psi \rangle_- =
\int d\eta d^3x \, \frac{1}{a^2}  \phi \psi \, ,
\eeq
respectively. Moreover, the operator $\epsilon$ is self-adjoint with any of these
inner products.

It is clear from the form of this Lagrangian that it is degenerate and thus describes
just one propagating field, not two. Indeed, the time derivatives of the fields enter in
the combination $(a^2/2)(\rho'-\chi'/a^2)^2$, which makes it clear that the system is
degenerate. There are many ways to unravel the dynamics of such a system, but the most
powerful, if not always the easiest method is via its Hamiltonian analysis. Thus, let us
rewrite it in the Hamiltonian form. To this end, we first rewrite the Lagrangian in the
form that makes it obvious what the canonical momenta are. Integrating by parts in the
terms containing second time derivatives, we write:
\ber
{\cal L}_{tens} &=& \frac{a^2}{2} \left(  \rho'^2 +\rho\Delta \rho \right) -
(\epsilon\rho - \im\rho') (\epsilon\chi+\im\chi') \nonumber \\ && {} + \frac{1}{2a^2}
\left[\chi'^2 + \chi \Delta \chi - 4 \im \cH \chi \epsilon \chi + \left( 4 \cH^2 + 8 \pi
G a^2 \kappa \right) \chi^2 \right] \, .
\eer
Now the conjugate momenta are:
\beq\label{tens-mom}
\pi_\rho = a^2 \rho' -\chi' +\im\epsilon\chi \, , \qquad \pi_\chi =
\frac{1}{a^2}\chi'-\rho' -\im\epsilon\rho \, .
\eeq
Thus, the following primary constraint holds:
\beq\label{constr-tens}
\phi_1:=\pi_\rho + a^2\pi_\chi + \im\epsilon(a^2\rho-\chi)=0 \, .
\eeq

The Hamiltonian is obtained as ${\cal H}_{\rm tens}=\pi_\rho \rho' +\pi_\chi \chi' -{\cal
L}_{\rm tens}+u\phi_1$, where $u$ is a Lagrange multiplier and $\phi$ is the primary
constraint. The Hamiltonian then must be written in terms of the momenta. Only one
momentum is needed for this, for which we choose $\pi_\rho$. We get:
\beq\label{H-tens}
{\cal H}_{\rm tens}=\frac{1}{2a^2} \pi_\rho^2 - \frac{1}{a^2} \pi_\rho \im\epsilon\chi
-\frac{a^2}{2} \rho\Delta\rho - \rho\Delta\chi +\frac{1}{2a^2} \left( 4\im\cH
\chi\epsilon\chi -4\cH^2\chi^2-8\pi G a^2 \kappa \chi^2 \right) + u\phi_1\, .
\eeq
As a check, one can verify that the Poisson brackets of this Hamiltonian with the phase
space variables $\pi_\rho$, $\pi_\chi$, $\rho$, and $\chi$ give rise precisely to the
original equations (\ref{tens-eq1}), (\ref{tens-eq2}) with the Lagrange multiplier
$u=\chi'/a^2$ as a consequence of $\chi'=\{H_{\rm tens},\chi\}=a^2 u$.

Now the condition that the constraint $\phi_1=0$ is preserved in time gives the secondary
constraint:
\beq
\phi_2:=\phi_1'= 2 a^2\cH (\pi_\chi +\im \epsilon\rho) + \{H_{\rm tens},\phi_1\}=0\, .
\eeq
We find:
\beq\label{tens-c2}
\phi_2:=a^2 \Delta \rho + \im\epsilon\pi_\rho -\cH\pi_\rho - \im\cH \epsilon\chi +
2\cH^2\chi + 4\pi G a^2 \kappa \chi = 0\, ,
\eeq
where we have used the relation $a^2(\pi_\chi+\im\epsilon\rho)=\im\epsilon\chi-\pi_\rho$
that follows from $\phi_1=0$. As a check, we note that (\ref{tens-c2}) is precisely the
difference of the two original equations (\ref{tens-eq1}), (\ref{tens-eq2}).

Now one can see that the Poisson bracket of $\phi_1$ with $\phi_2$ is a constant, and so
the constraints are second class. These constraints can be used to solve for $\pi_\chi$
and $\chi$ in terms of the phase space variables $\pi_\rho$ and $\rho$. After that, one
should substitute the result into (\ref{H-tens}) to obtain an effective Hamiltonian with
second class constraints solved for. In addition, one also has to compute the arising
Dirac bracket.

To solve (\ref{tens-c2}), we introduce:
\beq
m^2:=2\cH^2+4\pi G a^2 \kappa \, .
\eeq
We now get for $\chi$:
\beq\label{tens-chi}
\chi = \left( \im\cH\epsilon-m^2 \right)^{-1} \left[ a^2\Delta\rho + (\im \epsilon-\cH)
\pi_\rho \right]\, ,
\eeq
where the inverse of $\left( \im\cH\epsilon - m^2 \right)$ should be interpreted by
passing to the momentum space. There is no operator-ordering ambiguities in this formula
as all the operators only involve spatial derivatives and thus commute.

We now note that Hamiltonian (\ref{H-tens}) can be rewritten on the constraint surface
(\ref{tens-c2}) as
\beq
{\cal H}_{\rm tens}^\eff=\frac{1}{2a^2} \pi_\rho^2 -\frac{a^2}{2} \rho\Delta\rho -
\frac{\cH}{a^2} \chi (\pi_\rho -\im\epsilon\chi)\, ,
\eeq
where $\chi$ is to be substituted from (\ref{tens-chi}). The action in the Hamiltonian
form on the constraint surface now reads:
\beq
S^\eff_{tens}=\int d\eta \, d^3x \left[ \rho'\pi_\rho + \frac{\chi'}{a^2}\im\epsilon\chi
-\frac{\chi'}{a^2} \left( \pi_\rho + \im a^2\epsilon \rho \right) - {\cal H}^\eff_{\rm
tens} \right],
\eeq
where we have used the constraint (\ref{constr-tens}) to express $\pi_\chi$ in terms of
the other phase space variables. Now, the term
\beq
\frac{\chi'}{a^2}\im\epsilon\chi  - \frac{\cH}{a^2} \chi\im \epsilon\chi =
\left( \frac{\chi\im\epsilon\chi}{2a^2}\right)'
\eeq
here is a total time derivative and can be dropped. This leaves us with
\beq
S^\eff_{\rm tens}=\int d\eta \, d^3x \left[ \rho'\pi_\rho -\frac{\chi'}{a^2} \left(
\pi_\rho + \im a^2\epsilon \rho \right) -  \frac{1}{2a^2} \pi_\rho^2 + \frac{a^2}{2}
\rho\Delta\rho + \frac{\cH}{a^2} \chi \pi_\rho \right],
\eeq
where (\ref{tens-chi}) must be substituted. After this is done, and all the dust settles,
we get the following effective action in the Hamiltonian form:
\ber
S^\eff_{\rm tens} &=& \int d\eta \, d^3x \left( \rho' \frac{m^2 \pi_\rho}{m^2-\im
\cH\epsilon} -\frac{1}{2a^2}\pi_\rho^2 \left[ 1+ \left(\frac{\cH-\im\epsilon}{m^2-\im
\cH\epsilon}\right)'\right] \right. \nonumber \\
&& + \left. \Delta\rho \frac{\pi_\rho}{a} \left(\frac{a}{m^2-\im \cH\epsilon}\right)'
+\frac{a^2}{2}\rho\Delta\rho \left[ 1+ \frac{\im\epsilon}{a^2}  \left(\frac{a^2}{m^2-\im
\cH\epsilon}\right)'\right] \right)\, .
\eer
As before, the limit to GR is easily obtained by sending $m^2\to\infty$. Another simple
limit is that of passing to the Minkowski spacetime background. This is obtained by
setting $\cH=0$ and $a=1$ everywhere, as well as taking $m^2$ to be time-independent. One
again obtains an unmodified system, even for a finite $m^2$, which is consistent with the
earlier observations in the literature that the gravitational waves in the Minkowski
spacetime are unmodified. However, we see that tensor perturbations around an expanding
universe are modified quite non-trivially.

This action can be further rewritten in the usual form depending on generalized
velocities only by integrating out the momentum $\pi_\rho$. We get
\ber
S^\eff_{\rm tens} &=& \int d\eta \, d^3x \, \frac{a^2}{2} \left( \left[1+
\left(\frac{\cH-\im\epsilon}{m^2-\im \cH\epsilon}\right)'\right]^{-1} \left[ \rho'
\frac{m^2}{m^2-\im \cH\epsilon} +\frac{\Delta\rho}{a} \left(\frac{a}{m^2-\im
\cH\epsilon}\right)'\right]^2 \right. \nonumber \\ && + \left. \rho\Delta\rho \left[ 1+
\frac{\im\epsilon}{a^2}  \left(\frac{a^2}{m^2-\im \cH\epsilon}\right)'\right] \right)\, .
\label{tens-act-eff}
\eer
The resulting effective action is complex. The prescription for dealing with such complex
actions advocated in \cite{Krasnov:2009ik} was to require the metric to be real and take
the real part of the action. However, a different prescription may also be
possible.\footnote{One of the authors (KK) is grateful to Laurent Freidel for a
discussion that led to an alternative prescription.} Thus, it may be that the appropriate
prescription is to allow the action to be a complex, holomorphic function of the metric
$g_{\mu\nu}$, with the latter treated as a collection of complex variables. Then, the
condition that the imaginary part of the action vanishes gives a relation allowing to
express the imaginary part of the metric in terms of its real part. The real part of the
action can then be written as a functional of the real part of the metric. More work is
needed to decide which prescription for dealing with ``reality conditions'' for this
theory is appropriate. We leave an analysis of the tensor sector that depends on this
prescription to future research.

An alternative quick way to obtain the field equation for $\rho$ that stems from
(\ref{tens-act-eff}) is as follows. One notes that equation (\ref{tens-eq2}) can be
written as
\beq
\pi_\rho'-a^2\Delta\rho - \im\epsilon(\im\epsilon-\partial_\eta)\chi=0\, .
\eeq
On the other hand, from the definition (\ref{tens-mom}) of the momentum $\pi_\rho$, it
follows that
\beq
(\im\epsilon-\partial_\eta)\chi = \pi_\rho-a^2\rho' \, .
\eeq
This gives an equation that involves only $\rho$ and $\pi_\rho$:
\beq
\pi_\rho'-a^2\rho -\im\epsilon(\pi_\rho-a^2\rho')=0 \, .
\eeq
One can rewrite this as
\beq
(\partial_\eta-\im\epsilon)(\pi_\rho+\im a^2 \epsilon\rho) - 2\cH a^2 \im\epsilon\rho=0\,
.
\eeq
Using the expression for $\pi_\rho$ that was obtained above (when integrating it out), or
substituting $\chi$ into the definition (\ref{tens-mom}) of the momentum $\pi_\rho$ and
solving for $\pi_\rho$, we obtain the following compact expression for the combination
\beq
\pi_\rho+\im a^2 \epsilon\rho = a^2 \left[1+ \left(\frac{\cH-\im\epsilon}{m^2-\im
\cH\epsilon}\right)'\right]^{-1} (\partial_\eta+\im\epsilon) \frac{m^2
\rho}{m^2-\im\cH\epsilon}\, .
\eeq
As a result, we get the following differential equation for $\rho$:
\beq\label{eqn-rho}
(\partial_\eta-\im\epsilon)a^2 \left[1+ \left(\frac{\cH-\im\epsilon}{m^2-\im
\cH\epsilon}\right)'\right]^{-1} (\partial_\eta+\im\epsilon) \frac{m^2
\rho}{m^2-\im\cH\epsilon} - 2\cH a^2 \im\epsilon\rho=0\, .
\eeq
As a check, we note that the resulting equation takes the general-relativistic form
$(a^2\rho')'-a^2 \Delta\rho=0$ in the limit $m^2\to\infty$. We shall not analyze the
arising modified gravitational wave equations in this paper, leaving this to future work.

\end{document}